\documentclass[preprint]{elsarticle}

\usepackage{amsmath}
\usepackage{amssymb}

\title{Analysis, Design, and Generalization of Electrochemical Impedance Spectroscopy (EIS) Inversion Algorithms}
\author{Surya Effendy, Juhyun Song, and Martin Z. Bazant}

\begin{document}

\begin{center}
    \large{Analysis, Design, and Generalization of Electrochemical Impedance Spectroscopy (EIS) Inversion Algorithms}\\
    \normalsize{Surya Effendy$^a$, Juhyun Song$^a$, and Martin Z. Bazant$^{a\text{, }b\text{, }}$\footnote{Corresponding author. Email: bazant@mit.edu.}}\\
    \footnotesize{$^a$\textit{Massachusetts Institute of Technology, Department of Chemical Engineering, 25 Ames St., Cambridge, MA 02142, USA.}}\\
    \footnotesize{$^b$\textit{Massachusetts Institute of Technology, Department of Mathematics, 182 Memorial Dr., Cambridge, MA 02142, USA.}}
\end{center}

\section*{Abstract}
We introduce a framework for analyzing and designing EIS inversion algorithms. Our framework stems from the observation of four features common to well-defined EIS inversion algorithms, namely (1) the representation of unknown distributions, (2) the minimization of a metric of error to estimate parameters arising from the chosen representation, subject to constraints on (3) the complexity control parameters, and (4) a means for choosing optimal control parameter values. These features must be present to overcome the ill-posed nature of EIS inversion problems. We review three established EIS inversion algorithms to illustrate the pervasiveness of these features, and show the utility of the framework by resolving ambiguities concerning three more algorithms. Our framework is then used to design the generalized EIS inversion (gEISi) algorithm, which uses Gaussian basis function representation, modality control parameter, and cross-validation for choosing the optimal control parameter value. The gEISi algorithm is applicable to the generalized EIS inversion problem, which allows for a wider range of underlying models. We also considered the construction of credible intervals for distributions arising from the algorithm. The algorithm is able to accurately reproduce distributions which have been difficult to obtain using existing algorithms. It is provided gratis on the repository https://github.com/suryaeff/gEISi.git. 

\section{Introduction}\label{sec:Introduction}
Electrochemical impedance spectroscopy (EIS) is a central technique for analyzing electrochemical systems such as corroding surfaces \cite{Bonora96, Juttner90, Mansfeld90, Mansfeld95}, fuel cells \cite{Boukamp04, Fu15a, Fu15b, He09, Schichlein02}, batteries \cite{Andre11a, Andre11b, Deng13, Song13, Song14}, sensors \cite{Pejcic06}, solar cells \cite{Adachi06, Wang05}, and biological systems \cite{Lisdat08, Ruan02}. The application of a small oscillating current (or potential) results in a small oscillating potential (or current), which is then studied as a function of the angular frequency of the oscillation \cite{Barsoukov18, Lasia14, Orazem97}. The small magnitude of the input oscillation suppresses nonlinear behaviors in the frequency domain, allowing the system to be studied using relatively simple perturbation models \cite{Song19}. The ratio of the oscillating potential to the oscillating current in frequency domain, i.e., the impedance, often exhibits features which are easily correlated with the parameters of the model. 

The success of EIS data analysis is contingent on the ability to express the system as a quantitative physical model, but it is often the case that the system is not sufficiently well-understood to allow such a model to be written. In such cases, the distribution of relaxation time (DRT) problem may be solved, wherein the underlying model is assumed to be a series of relaxation processes with a distribution of characteristic timescales \cite{Boukamp15, Ciucci15, Florsch14, Hershkovitz11a, Hershkovitz11b, Horlin98, Saccoccio14, Han15}. The DRT problem is a linear Fredholm integral of the first kind, which can be solved via Fourier, Laplace, or Mellin transform, although questions concerning the existence and uniqueness of the resulting solution inevitably arise \cite{Tricomi85, Whittaker96}. Similar inverse problems have a long history in statistical mechanics, where microscopic distributions, such as the density of states or the partition function, are directly inverted from thermodynamic data \cite{Bazant01}. Resolution of the DRT problem yields the distribution of characteristic timescales, which can be used to identify the number, size, and average timescale of physical processes within the system \cite{Boukamp15, Ciucci15, Hershkovitz11a, Horlin98}, and in some cases, assess the propriety of the assumed relaxation process \cite{Boukamp15, Hershkovitz11a, Horlin98}. The DRT problem also allows resolution of overlapping features in the impedance spectra, whose visual identification is often challenging. The DRT model has found increasing acceptance over the past decade, as evidenced by recent works on batteries \cite{Ciucci15, Klotz12, Li18, Saccoccio14, Schmidt13, Han15}, fuel cells \cite{Boukamp15, Hershkovitz11a, Schindler16, Sumi17, Zhang16}, and geophysics \cite{Florsch14}.

This acceptance may be partly attributed to the relative maturity of DRT inversion algorithms, i.e., algorithms used to obtain the DRT, which span a range of approaches such as Fourier transform \cite{Boukamp15, Schichlein02}, construction of L-curves \cite{Florsch14}, genetic algorithm \cite{Hershkovitz11a, Hershkovitz11b}, maximum entropy method \cite{Horlin98}, various least-squares methods \cite{Macdonald95, Tuncer06}, various regularization methods \cite{Ciucci15, Saccoccio14, Song18, Tuncer06, Zhang16, Zic20}, and Monte Carlo method \cite{Tuncer01}. Generic inversion toolboxes \cite{Hansen94} have also been used to solve the DRT problem \cite{Dion99}. This impressive breadth of approaches poses an interesting challenge concerning the classification and design of DRT inversion algorithms. As noted by Saccoccio, Han, Chen, and Ciucci, there is a need for ``speed and accuracy benchmarks for the various methods" \cite{Saccoccio14}, so as to allow one to characterize the optimality of inversion algorithms, and choose an appropriate algorithm given the nature of the problem and the availability of computational resource. Concurrently, there is a need for a framework which can be used to analyze the optimality of inversion algorithms, so as to allow the design of an appropriate algorithm given the nature of the problem and the availability of computational resource. 

We consider one such framework, founded upon the observation that there are four features common to all well-posed EIS inversion algorithms. These features can be broadly stated as (1) the representation of unknown distributions, (2) the minimization of a metric of error to estimate the parameters arising from the chosen representation, subject to constraints on (3) the complexity control parameters, and (4) a means for choosing optimal control parameter values. For brevity, we will refer to these four features as the representation, the interior problem, the complexity control parameter, and the exterior problem. We illustrate the prevalence of the framework through a review of three established DRT inversion algorithms, and show the utility of the framework by resolving ambiguities and difficulties concerning three more DRT inversion algorithms. We then introduce the generalized EIS inversion problem, which expands upon the concept of DRT to allow for a wider range of underlying models, design the corresponding generalized EIS inversion (gEISi) algorithm, and consider the construction of credible intervals for the parameters arising from the algorithm. The gEISi algorithm is able to reproduce distributions which exhibit a wide range of smoothness, which has been difficult to perform using existing EIS inversion algorithms. We also present validation against synthetic data generated from the distributed Randles circuit, which is impossible to analyze using existing EIS inversion algorithms. The application of the gEISi algorithm to experimental data modelled by the distributed Randles circuit will be the subject of a future companion paper \cite{effendy21}.

\section{Overview of the Concept}\label{sec:Overview of the Concept}
We begin with the nomenclature of terms used in the present work. The term \textit{EIS inversion problem} refers to a general class of problems in which distributions are inferred from the impedance spectra, assuming some simplified underlying physics. The classical example of an EIS inversion problem is the \textit{DRT problem}, which has been introduced in Section \ref{sec:Introduction}:
\begin{equation}\label{eq:DRT Problem without Transform}
    \hat{Z}(\omega;R_\infty,\gamma(\tau))=\int_0^\infty{\frac{\gamma(\tau)}{1+i\omega\tau}d\tau}+R_\infty
\end{equation}
The DRT problem uses the Debye model, which assumes an underlying relaxation process \cite{Barsoukov18}. Note that $\hat{Z}(\omega;R_\infty,\gamma(\tau))$ implies that the predicted impedance ($\hat{Z}$) has the angular frequency ($\omega$) as independent variable, and the Ohmic resistance ($R_\infty$) as well as the DRT ($\gamma(\tau)$) as model parameters. The unit of $\gamma(\tau)$ is $\Omega\text{/s}$, indicating that $\gamma(\tau)$ is not a probability distribution. Data from the literature suggests that the domain of $\gamma(\tau)$ can spread over several orders of magnitude \cite{Amin07, Han14}. It is thus convenient to perform the substitution $v\equiv\ln(\tau)$:
\begin{equation}\label{eq:DRT Problem with Transform}
    \hat{Z}(\omega;R_\infty,G(v))=\int_{-\infty}^\infty{\frac{G(v)}{1+i\omega\exp(v)}dv}+R_\infty
\end{equation}
Here $G(v)=\tau \gamma(\tau)$ is the DRT on a logarithmic scale. The objective of the DRT problem is to estimate $G(v)$ given a vector of measured impedance values $\textbf{Z}$, with $Z(\omega_j)$ as the $j$-th component of $\textbf{Z}$.

Florsch, Revil, and Camerlynck proposed a variation to the EIS inversion problem, wherein the Debye model is replaced with the Havrilliak-Negami model \cite{Florsch14}:
\begin{equation}\label{eq:Florsch Problem}
    \hat{Z}(\omega;R_\infty,G(v))=\int_{-\infty}^\infty{\frac{G(v)}{\left[1+(i\omega\exp(v))^\alpha\right]^\beta}}dv+R_\infty
\end{equation}
Here $\alpha$ and $\beta$ are constants which are determined prior to solving the EIS inversion problem. In particular, Florsch considered the case whereby $\alpha=0.5$ and $\beta=1$, which corresponds to an infinite series of parallel diffusive-capacitive processes. The motivation for using the Havrilliak-Negami model is simple, yet significant: if the correct model is chosen, the distribution $G(v)$ should exhibit low complexity. This supposition is used to conclude that the impedance data obtained from their archaeological sample can be best explained by the polarization of an electronically conductive sample.

Yet another variation to the EIS inversion problem is the \textit{distribution of diffusion time (DDT) problem} \cite{Song18}, which is appropriate for cases whereby prior knowledge suggests that the underlying physics is better explained by a diffusive process. The model is chosen based on the nature of the diffusive process:
\begin{equation}\label{eq:DDT Problem}
    \hat{Y}(\omega;G(v))=\int_{-\infty}^\infty{G(v)\hat{y}(\omega,v)dv}
\end{equation}
Here $\hat{Y}$ is the predicted admittance, and $\hat{y}$ describes the underlying diffusion process. Song and Bazant provides a list of expressions for $\hat{y}$ based on the boundary conditions and the symmetry of the system under consideration \cite{Song18}. The integral is performed over the admittance rather than the impedance, so as to reflect the parallel nature of diffusion normal to the surface of electrodes. The DDT problem was solved for the impedance spectra of a silicon-nanowire Li-ion battery anode \cite{Ruffo09}, yielding a distribution of nanowire radii, which was subsequently confirmed against scanning electron microscopy image analysis. 

The works of Florsch, Revil, and Camerlynck \cite{Florsch14} and Song and Bazant \cite{Song18} demonstrate that prior knowledge can be used to construct the underlying physics of an EIS inversion problem, such that the resulting distribution $G(v)$ corresponds to an observable, as opposed to a mathematical construct whose utility is restricted to the identification of the number, size, and average timescale of underlying processes \cite{Boukamp15, Hershkovitz11a, Horlin98}. This motivates the idea of a \textit{generalized EIS inversion problem}, in which the form of the underlying physics is not specified a priori, and the corresponding \textit{generalized EIS inversion algorithm}, which solves the problem independently from the underlying physics. We will explore this concept in greater detail in Section \ref{sec:Design of Algorithm}.

The remainder of this section, which discusses the proposed framework, and Section \ref{sec:Analysis}, which reviews select algorithms, will focus on the DRT problem, given its prevalence in the literature. However, the concepts outlined therein are also applicable to the more general class of EIS inversion problem.

\subsection{Framework}\label{subsec:Framework}
We claim that all well-posed DRT inversion algorithms possess four common features, namely, representation, interior problem, complexity control parameter, and exterior problem. The representation approximates the unknown distribution with a discrete or continuous set of basis functions. The most common representation is the sum of triangular basis functions:
\begin{align}
    G(v;\textbf{g})&\equiv{\sum_{m=1}^M{g_m\Delta_m(v)}}\label{eq:Triangular Basis Representation}\\
    \Delta_m(v)&\equiv\begin{cases}\label{eq:Triangular Basis}
    \frac{v-v_{m-1}}{v_m-v_{m-1}}\quad\text{if}\quad{v\in[v_{m-1},v_m]}\\
    \frac{v_{m+1}-v}{v_{m+1}-v_m}\quad\text{if}\quad{v\in[v_m,v_{m+1}]}\\
    0\quad\text{otherwise}
    \end{cases}
\end{align}
Here $\textbf{g}$ is the vector of basis function weights with $g_m$ as the $m$-th component of $\textbf{g}$, $m$ is the index of the basis functions, $M$ is the total number of basis functions, and $\Delta_m(v)$ is the basis function. This representation is functionally identical to linear interpolation, and possesses variations according to the treatment of the end-points \cite{Macdonald95}.

\begin{figure*}[h!]
    \centering
    \includegraphics[scale=0.6]{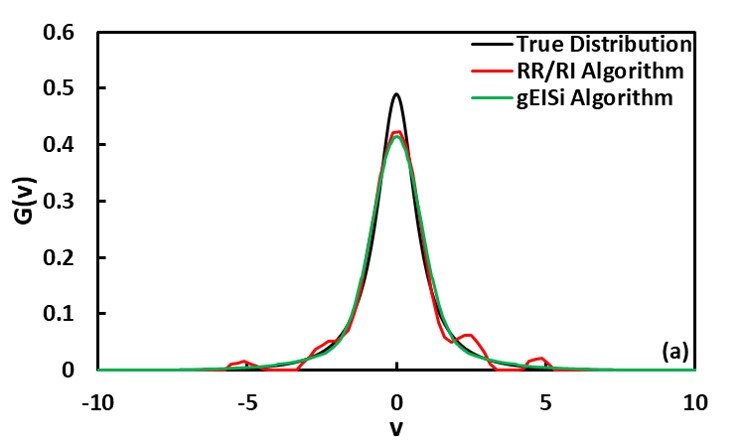}
    \includegraphics[scale=0.6]{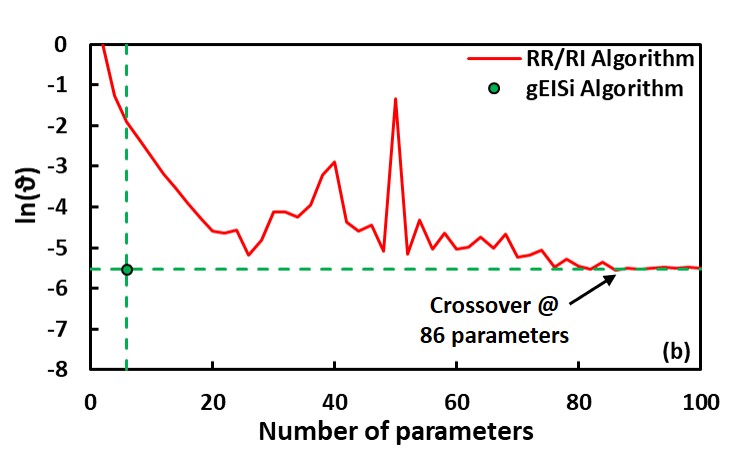}
    \caption{(a) True and inverted distributions corresponding to a unimodal Cole-Cole distribution, obtained using the RR/RI algorithm and the gEISi algorithm. (b) The accuracy of the representation as a function of the number of parameters used in the RR/RI algorithm. The gEISi algorithm returns a DRT containing 6 parameters.}
    \label{fig:Parameter Reduction}
\end{figure*}

It is useful to distinguish between \textit{fixed-mesh} and \textit{floating-mesh} representations. Floating-mesh representations incorporate the mesh points as parameters in the basis function expansion, while fixed-mesh representations do not. For example, Equation (\ref{eq:Triangular Basis Representation}) is a fixed-mesh representation, since it is only parameterized by the weights of the basis function ($g_m$), and not by the location of the mesh points ($v_m$). This distinction is significant, because floating-mesh representations tend to approximate distributions using a smaller number of parameters. To illustrate this point, consider the true underlying distribution shown in Figure \ref{fig:Parameter Reduction} (a), which is unimodal Cole-Cole centered at $t=0$. We collect 10 noisy impedance measurements per decade between $\omega_1=10^{-4}$ and $\omega_J=10^4$, and define the accuracy of a representation as:
\begin{equation}
    \vartheta\left[G(v)\right]\equiv\int_{-\infty}^\infty[G(v)-G_\text{true}(v)]^2dv
\end{equation}
Here $G_\text{true}(v)$ is the true underlying distribution. We invert the underlying distribution using the RR/RI algorithm introduced in Section \ref{subsec:RR/RI Algorithm}, which employs a fixed-mesh representation, and the gEISi algorithm introduced in Section \ref{subsec:gEISi Algorithm}, which employs a floating-mesh representation. The mesh boundaries of the RR/RI algorithm are determined from the upper and lower limits of the measurement angular frequencies \cite{Ciucci15, Florsch14, Macdonald95, Saccoccio14, Song18, Tuncer06, Han15}:
\begin{align}\label{eq:Mesh Boundaries}
    v_0&=-\ln(\omega_J)\\
    v_{M+1}&=-\ln(\omega_1)
\end{align}
We then adjust the number of parameters used in the RR/RI algorithm to match the accuracy of the gEISi algorithm.

The unimodal Cole-Cole distribution is accurately reproduced by both algorithms, but the RR/RI algorithm requires 86 parameters to match the accuracy of the gEISi algorithm, which only requires 6 parameters. This order-of-magnitude reduction in the number of parameters does not translate to a reduction in computational time, which rather increases by an order of magnitude. The relative efficiency of the RR/RI algorithm \cite{Saccoccio14}, as well as the majority of algorithms employing the fixed-mesh representation \cite{Ciucci15, Florsch14, Song18, Han15}, arises from the use of near-analytical linearly constrained quadratic programming. In contrast, the floating-mesh representation almost inevitably requires stochastic programming to solve \cite{Hershkovitz11a, Hershkovitz11b}, which can be computationally expensive. Nevertheless, this reduction in the number of parameters will be significant in context of the generalized EIS inversion problem, which is discussed in Section \ref{subsec:Generalized EIS Inversion Problem}. 

The representation of the unknown distribution leads to a set of unknown parameter values, which are estimated via the interior problem, subject to a constraint on the complexity control parameter. The interior problem shows the least variation among the four common features, and is typically the minimization of a misfit. The idea of estimating parameter values subject to a constraint on the control parameter will be demonstrated in a subsequent paragraph.  

The third common feature is the complexity control parameter. Implicit in the works of Florsch, Revil, and Camerlynck \cite{Florsch14} and Song and Bazant \cite{Song18}, which have been discussed in Section \ref{sec:Overview of the Concept}, is the idea that the application of a proper underlying model tends to result in a distribution with low complexity. As a corollary, if two different underlying models result in two distributions with different complexities, the underlying model yielding the distribution of lower complexity should be favored. It is important to recognize outright that this is a restatement of Occam's razor \cite{Balasubramanian97, Blumer87, Domingos99}, i.e., among competing hypotheses, all else being equal, favor that which requires the least assumptions. We will explore the relationship between complexity and Occam's razor in greater depth in Section \ref{sec:Design of Algorithm}. The existence of complexity is also recognized by Hershkovitz, Tomer, Baltianski, and Tsur, who defines complexity as the number of parameters needed to approximate a DRT \cite{Hershkovitz11a, Hershkovitz11b}.

Intuitively, we grasp that a highly fluctuating distribution is complex, while a smoothly varying one is not. The idea of complexity is central to the DRT problem, because the ill-posed nature of the problem is in effect resolved by limiting the complexity of the distribution. Consider the hierarchical Bayesian algorithm of Ciucci and Chen \cite{Ciucci15}. For an exponential hyperprior, it can be shown that the interior problem takes the form:
\begin{equation}\label{eq:Interior Problem Ciucci}
    \textbf{g}^*\equiv\underset{\textbf{g}}{\text{argmin}}\enskip\underbrace{\left\lVert{\textbf{Z}-\hat{\textbf{Z}}}\right\rVert_2^2}_{\text{misfit}}+\underbrace{\left\lVert{\ln\left[(\textbf{L}_1\textbf{g})^2+\xi\right]}\right\rVert_1}_{\text{penalty}}
\end{equation}
Here $\textbf{L}_1$ is the first derivative operator, $\xi$ is a parameter describing the width of the hyperprior, and $\left\lVert{\cdot}\right\rVert_1$ and $\left\lVert{\cdot}\right\rVert_2$ are the first and second norm operators, respectively. This interior problem may be thought of as the minimization of model misfit subject to a penalty on the first derivative of $\textbf{g}$. The parameter $\xi$ controls the complexity of the resulting DRT. When $\xi$ becomes large, the penalty becomes insensitive to changes in the first derivative, yielding a highly oscillatory distribution. In contrast, when $\xi$ becomes small, due to the singular nature of the $\ln[\cdot]$ operator, $\textbf{L}_1\textbf{g}$ tends to vanish everywhere, yielding a more uniform distribution. 

Hereon, we will define the complexity control parameter as a quantitative constraint on the complicated behavior of a DRT. In the hierarchical Bayesian algorithm, the control parameter is $\xi$. It is useful to distinguish between the complexity control parameter and the complexity, the latter being a description of the complicated behavior. In the context of the hierarchical Bayesian algorithm, the complexity is the first derivative of the distribution, i.e., $(\textbf{L}_1\textbf{g})^2$. 

The exterior problem selects an appropriate value for the control parameter. The exterior problem can be quantitative, as is the case for most EIS inversion algorithms, but it can also be qualitative, as was done by Boukamp, who uses common sense for the ``adjustment of the [parameters of the] window function, which should lead to an `acceptable' DRT" \cite{Boukamp15}. Note that the parameters of the window function are the control parameters of Boukamp's DRT inversion algorithm. In contrast to the the representation, which is largely dominated by triangular basis functions, or the interior problem, which is typically the minimization of a misfit, the exterior problem exhibits great variation in form. We will discuss these in greater detail in subsequent sections.

\begin{figure}
    \centering
    \includegraphics[scale=0.5]{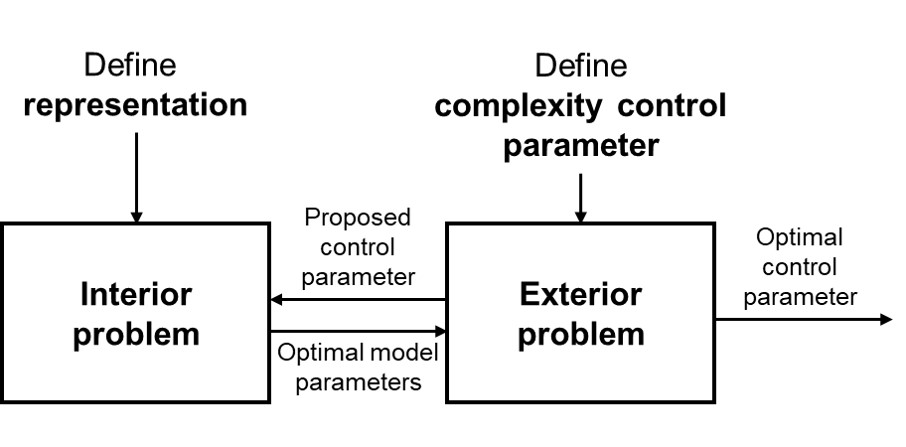}
    \caption{The common features of well-posed EIS inversion algorithms. After defining the representation and the complexity control parameter, one can set up an interior problem which estimates the unknown parameters subject to a constraint on the control parameter, and an exterior problem which determines an appropriate value for the control parameter.}
    \label{fig:RICE Framework}
\end{figure}

The relationship between the four common features described in this section has been summarized in Figure \ref{fig:RICE Framework}. EIS inversion algorithms can be constructed by defining the four common features, which act as standardized modules in a program. We will now demonstrate the prevalence of the four common features in the literature.

\section{Sample Algorithms}\label{sec:Sample Algorithms}
In this section, we classify three distinct EIS inversion algorithms according to the four common features. The first algorithm, taken from Saccoccio, Han, Chen, and Ciucci, consists of ridge regression coupled to a real-imaginary cross-validation routine \cite{Saccoccio14, Han15}. The work of Saccoccio, Han, Chen, and Ciucci is notable for its thorough investigation of the real-imaginary cross-validation routine, which provides compelling evidence for the effectiveness of the routine in cases whereby the error in the real part of the impedance is independent of the error in the imaginary part of the impedance. Hereon, we will refer to this algorithm as the ridge regression / real-imaginary cross-validation (RR/RI) algorithm. The second algorithm is the least-squares vector minimization (LEVM) algorithm of Macdonald \cite{Macdonald95, Tuncer06}. This algorithm possesses an implicit exterior problem, which will become apparent through a comprehensive analysis of the surrounding literature. The third algorithm, adapted from Hershkovitz, Tomer, Baltianski, and Tsur, provides an example of the four common features in a DRT inversion algorithm which uses a floating-mesh representation \cite{Hershkovitz11a, Hershkovitz11b}. In deference to the work of Hershkovitz, Tomer, Baltianski, and Tsur, we will refer to this algorithm as the impedance spectroscopy analysis using genetic programming (ISGP) algorithm. 

The algorithms considered in this and subsequent sections are not presented in their original form. To improve ease of understanding, simplifications have been made to parts of these algorithms, and various symbols have been changed to align with the present work. Unless stated otherwise, these changes are superficial, and should not alter the behavior of the algorithms substantially. 

\subsection{RR/RI Algorithm}\label{subsec:RR/RI Algorithm}
The work of Saccoccio, Han, Chen, and Ciucci provides the quintessential example of the four common features. The DRT problem solved by Saccoccio, Han, Chen, and Ciucci is a simplified form of Equation (\ref{eq:DRT Problem with Transform}), where the Ohmic resistance has been manually eliminated from the experimental data.
\begin{equation}\label{eq:Simplified DRT Problem}
    \hat{Z}(\omega;G(v))=\int_{-\infty}^\infty{\frac{G(v)}{1+i\omega\exp(v)}dv}
\end{equation}
The chosen representation is the sum of triangular basis functions, i.e., Equation (\ref{eq:Triangular Basis Representation}), which introduces $\textbf{g}$ as unknown parameter values:
\begin{equation}
\hat{Z}(\omega;\textbf{g})={\int_{-\infty}^\infty\frac{G(v;\textbf{g})}{1+i\omega\exp(v)}dv}
\end{equation}

The interior problem is composed of two optimization subroutines:
\begin{align}
    \textbf{g}^\mathbb{I}&\equiv\underset{\textbf{g}}{\text{argmin}}\enskip\left\lVert{\text{im}[\textbf{Z}-\hat{\textbf{Z}}]}\right\rVert_2^2+\lambda\left\lVert{\textbf{L}_1\textbf{g}}\right\rVert_2^2\label{eq:Imaginary Regularization Saccoccio}\\
    \textbf{g}^\mathbb{R}&\equiv\underset{\textbf{g}}{\text{argmin}}\enskip\left\lVert{\text{re}[\textbf{Z}-\hat{\textbf{Z}}]}\right\rVert_2^2+\lambda\left\lVert{\textbf{L}_1\textbf{g}}\right\rVert_2^2\label{eq:Real Regularization Saccoccio}
\end{align}
Here $\lambda$ is the regularization parameter, and re[$\cdot$] and im[$\cdot$] are the real and imaginary operators, respectively. These equations are similar to Equation (\ref{eq:Interior Problem Ciucci}), in that they involve the minimization of misfit subject to a penalty on the first derivative of $\textbf{g}$, but differ in the use of only the imaginary (Equation (\ref{eq:Imaginary Regularization Saccoccio})) and the real (Equation (\ref{eq:Real Regularization Saccoccio})) parts of the data set. The superscripts $\mathbb{I}$ and $\mathbb{R}$ denote that the parameters are obtained using only the imaginary and real parts of the data set, respectively. 

The complexity control parameter is the inverse of the regularization parameter ($1/\lambda$), and the complexity is the first derivative of the DRT. This complexity control parameter makes sense, because increasing $\lambda$ causes the penalty on the square of the first derivative to grow, resulting in smoother probability distributions.

The exterior problem is the real-imaginary cross-validation routine, wherein $\mathbf{g}^\mathbb{I}$ is used to predict the real part of the data set, and $\mathbf{g}^\mathbb{R}$ is used to predict the imaginary part of the data set:
\begin{equation}\label{eq:Exterior Problem Saccoccio}
    \lambda^*\equiv\underset{\lambda}{\text{argmin}}\enskip\left\lVert{\text{re}\left[\textbf{Z}-\hat{\textbf{Z}}(\textbf{g}^\mathbb{I})\right]}\right\rVert_2^2+\left\lVert{\text{im}\left[\textbf{Z}-\hat{\textbf{Z}}(\textbf{g}^\mathbb{R})\right]}\right\rVert_2^2
\end{equation}
The interior problem is embedded within the exterior problem, because $\mathbf{g}^\mathbb{I}$ and $\mathbf{g}^\mathbb{R}$ are both functions of $\lambda$. When excessively low control parameter values are assumed, the inferred DRT is over-smoothed, causing the cross-validation error to exhibit a large bias component, and when excessively high control parameter values are assumed, the inferred DRT is under-smoothed, causing the cross-validation error to exhibit a large variance component. The trade-off between bias and variance is discussed in greater detail by Saccoccio, Han, Chen, and Ciucci \cite{Saccoccio14}.

\subsection{LEVM Algorithm}\label{subsec:LEVM Algorithm}
The LEVM algorithm allows for several representations. The present work will focus on the continuous variable representation, which has been reported to outperform its discrete and fixed counterparts. This representation takes the form:
\begin{align}
    G(v;\textbf{g},\textbf{v})&\equiv{\sum_{m=1}^M{g_m\Delta_m(v;\textbf{v})}}\label{eq:Triangular Basis Representation Macdonald}\\
    \Delta_m(v;\textbf{v})&\equiv\begin{cases}
    \frac{v-v_{m-1}}{v_{m+1}-v_{m-1}}\quad\text{if}\quad{v\in[v_{m-1},v_{m+1}]}\\
    \frac{v_{m+1}-v}{v_{m+1}-v_m}\quad\text{if}\quad{v\in[v_m,v_{m+1}]}\\
    0\quad\text{otherwise}
    \end{cases}
\end{align}
The location of the mesh points ($\textbf{v}$) parameterizes the model, making this a floating-mesh representation. The interior problem is the minimization of misfit:
\begin{equation}\label{eq:Interior Problem Macdonald}
    \textbf{g}^*,\textbf{v}^*\equiv\underset{\textbf{g},\textbf{v}}{\text{argmin}}\left\lVert{\textbf{Z}-\hat{\textbf{Z}}}\right\rVert_2^2
\end{equation}

The representation and the interior problem of the LEVM algorithm are well-characterized, but there are no apparent control parameter and exterior problem. However, we can infer the existence of the control parameter from the discussion presented by Macdonald on the inversion of water data \cite{Macdonald95}. Macdonald has, in effect, taken the total number of mesh points ($M$) as the complexity control parameter, which is sensible, because a more complicated DRT would require more mesh points to approximate. Conversely, through a priori specification of $M$, Macdonald constrains how complicated the DRT can be. This observation is supported by the analysis of the output of the interior problem, in which $M$ is increased to improve the normalized misfit\footnote{Equation (\ref{eq:Exterior Problem 1 Macdonald}) is reconstructed from the description of the normalized misfit provided in Appendix B of reference \cite{Macdonald95}. This exterior problem assumes that the measurement error of the real and the imaginary parts of the impedance are completely correlated, thus yielding J independent measurements.}:
\begin{equation}\label{eq:Exterior Problem 1 Macdonald}
    M^*\equiv\underset{M}{\text{argmin}}\enskip\frac{\left\lVert{\textbf{Z}-\hat{\textbf{Z}}(\textbf{g}^*,\textbf{v}^*)}\right\rVert_2^2}{J-2M-1}
\end{equation}
As per the case of Equation (\ref{eq:Exterior Problem Saccoccio}), the interior problem is embedded within the exterior problem, as $\textbf{g}^*$ and $\textbf{v}^*$ are both functions of $M$. This exterior problem implies that each additional mesh point results in diminishing improvement in misfit, and so the number of mesh points should be balanced against the degree of freedom left in the data set ($J-2M-1$). However, in the subsequent discussion on the inversion of n-pentanol and glycerol data, it became evident that Equation (\ref{eq:Exterior Problem 1 Macdonald}) is insufficient. For the glycerol data, with a total of 10 mesh points, ``the [normalized misfit] was about 0.0094, and the estimated relative standard deviations of the [$\textbf{g}$] were all below 0.1 and mostly below 0.06", but when the number of mesh points is increased to 11, ``the relative standard deviations of 6 of the 11 [$\textbf{g}$] were then above 0.1, with one above 0.34", leading Macdonald to reject the solution containing 11 mesh points despite a continued decrease in normalized misfit. Macdonald thus implicitly recognizes the statistical significance of $\textbf{g}$ as constraints in the exterior problem:
\begin{equation}\label{eq:Exterior Problem 2 Macdonald}
    \overline{\sigma}(\textbf{g})\leq{0.34}
\end{equation}
Here $\overline{\sigma}[\cdot]$ is the relative standard deviation operator. Equation (\ref{eq:Exterior Problem 1 Macdonald}) and Equation (\ref{eq:Exterior Problem 2 Macdonald}) constitute the full exterior problem. 

\subsection{ISGP Algorithm}\label{subsec:ISGP Algorithm}
The ISGP algorithm employs a flexible representation composed of multiple basis functions \cite{Hershkovitz11a, Hershkovitz11b}:
\begin{equation}
    G(v;\{G_l\},\{\textbf{g}_l\})\equiv\sum_{l=1}^L{G_l(v;\textbf{g}_l)}
\end{equation}
The $l^\text{th}$ basis function $\left(G_l(v;\textbf{g}_l)\right)$ may be Gaussian, Lorentzian, hyperbolic secant, Kirkwood-Fuoss or Cole-Cole, each having three parameters, or Pearson VII or Havrilliak-Negami, each having four parameters, or the five-parameter pseudo-Voigt distribution. The DRT is thus parameterized by the type of the basis function as well as the parameters of those basis functions. To illustrate the indexing, consider the case whereby $L=2$:
\begin{align}
    G(v;G_1,G_2,\textbf{g}_1,\textbf{g}_2)&=G_1(v;\textbf{g}_1)+G_2(v;\textbf{g}_2)\\
    G_1(v;\textbf{g}_1)&=\frac{g_{1,1}}{\sqrt{2\pi}g_{3,1}}\exp\left[-\frac{(v-g_{2,1})^2}{2g_{3,1}^2}\right]\\
    G_2(v;\textbf{g}_2)&=\frac{g_{1,2}}{2\pi}\frac{\sin([1-g_{3,2}]\pi)}{\cosh(g_{3,2}[v-g_{2,2}])-\cos([1-g_{3,2}]\pi)}
\end{align}
Thus the DRT is represented by the sum of a Gaussian distribution ($G_1$) with mass $g_{1,1}$, mean $g_{2,1}$ and standard deviation $g_{3,1}$, and a Cole-Cole distribution with mass $g_{1,2}$, mean $g_{2,2}$ and depression $g_{3,2}$. In principle, the large number of possible basis functions allows the form of the underlying model to be detected, but the success of such an approach has not been demonstrated.

The ISGP requires as input two sets of impedance data ($\textbf{Z}_1$ and $\textbf{Z}_2$). The first data set is used to solve the interior problem, which is the minimization of misfit:
\begin{equation}\label{eq:Interior Problem Hershkovitz}
    \{G_l^*\},\{\textbf{g}_l^*\}\equiv\underset{\{G_l\},\{\textbf{g}_l\}}{\text{argmin}}\enskip\left\lVert{\textbf{Z}_1-\hat{\textbf{Z}}}\right\rVert_2^2
\end{equation}
This interior problem is not compatible with gradient-based optimization methods, since it considers the type of basis function ($\{G_l\}$) as a decision variable, thus motivating the use of genetic programming.

The ISGP uses the total number of parameters needed to represent the DRT ($M$) as the complexity control parameter, which is in line with the intuition that complicated distributions possess more features, and thus require more parameters to approximate. Likewise, by constraining $M$, it becomes possible to limit how complicated the DRT can be.

Hershkovitz, Tomer, Baltianski, and Tsur \cite{Hershkovitz11b} reported the following exterior problem:
\begin{align}\label{eq:Exterior Problem 1 Hershkovitz}
    M^*&\equiv\underset{M}{\text{argmin}}\enskip\frac{0.8\,\epsilon_1^2+0.2\,\epsilon_2^2}{1+\exp(M-M_H)}\\
    \epsilon_1^2&=\left\lVert{\textbf{Z}_1-\hat{\textbf{Z}}(\{G_l^*\},\{\textbf{g}_l^*\})}\right\rVert_2^2\\
    \epsilon_2^2&=\left\lVert{\textbf{Z}_2-\hat{\textbf{Z}}(\{G_l^*\},\{\textbf{g}_l^*\})}\right\rVert_2^2
\end{align}
Here $M_H$ is the recommended maximum number of parameters in the representation of the DRT, which is specified by the user, $\epsilon_1^2$ is the misfit arising from the interior problem, and $\epsilon_2^2$ is the validation misfit. This exterior problem is likely erroneous, because as $M$ increases, the objective function in Equation (\ref{eq:Exterior Problem 1 Hershkovitz}) decreases. This objective function thus favors models with $M$ in excess of $M_H$, which contradicts the description of $M_H$ as the recommended maximum number of parameters. We assume that this is a straightforward case of mistaking a multiplication for a division, and the exterior problem is supposed to be:
\begin{equation}\label{eq:Exterior Problem 2 Hershkovitz}
    M^*\equiv\underset{M}{\text{argmin}}\enskip\left[0.8\,\epsilon_1^2+0.2\,\epsilon_2^2\right]\times\left[1+\exp(M-M_H)\right]
\end{equation}
In Section \ref{sec:Design of Algorithm}, we will use the ISGP algorithm with Equation (\ref{eq:Exterior Problem 2 Hershkovitz}) as the exterior problem.

\section{Analysis}\label{sec:Analysis}
The examples discussed in Section \ref{sec:Sample Algorithms} illustrate the prevalence of the four common features. In this section, we consider the utility of the four common features as a tool for analyzing and supplementing known DRT inversion algorithms. The Fourier transform algorithm of Schichlein, M\"{u}ller, Voigts, Kr\"{u}gel, and Ivers-Tiff\'{e}e \cite{Schichlein02} and Boukamp \cite{Boukamp15} lacks parameters which describe the width of the window function. An exterior problem which identifies an optimal width is constructed. The L-curve algorithm of Florsch, Revil, and Camerlynck \cite{Florsch14} lacks a parameter which describes the rotation of the L-curve, which is needed to complete the exterior problem. Manipulation of the exterior problem yields equivalent forms which carry statistical meaning, allowing the rotation parameter to be determined. The maximum entropy algorithm of H\"{o}rlin is well-defined \cite{Horlin98}, but does not seem to work well for certain smooth distributions, which is unusual for a maximum entropy method. We will use the four common features to analyze this unexpected behavior.
 
\subsection{Fourier Transform Algorithm}\label{subsec:Fourier Transform Algorithm}
The Fourier transform algorithm of Schichlein, M\"{u}ller, Voigts, Kr\"{u}gel, and Ivers-Tiff\'{e}e \cite{Schichlein02} and the subsequent modified version published by Boukamp \cite{Boukamp15} are founded upon the deconvolution theorem, which states that convolution integrals are multiplications in the frequency domain:
\begin{equation}\label{eq:Deconvolution Theorem}
    h(x)=\int_{-\infty}^\infty{f(y-x)g(y)}dy\enskip\text{if and only if}\enskip\mathcal{F}(h)=\mathcal{F}(f)\mathcal{F}(g)
\end{equation}
Here $\mathcal{F}[\cdot]$ is the Fourier transform, which is defined as:
\begin{equation}
    \mathcal{F}[h(x)]\equiv\int_{-\infty}^\infty{h(x)\exp(-2\pi{ixs}})ds
\end{equation}
Note that $f$, $g$, $h$, $x$ and $y$ are dummy variables, and will not be used in the remainder of this section. 

Equation (\ref{eq:Simplified DRT Problem}) is transformed to a convolution integral by performing the substitution $u=-\ln(\omega)$:
\begin{equation}\label{eq:DRT convolution}
    \hat{Z}(u;G(v))=\int_{-\infty}^\infty{\frac{G(v)}{1+i\exp(v-u)}dv}
\end{equation}
The real and imaginary parts of $\hat{Z}(u;G(v))$ are linked by the Kramers-Kronig relations. It is thus sufficient to consider the imaginary part of the impedance:
\begin{equation}
    \hat{I}\equiv{-2\,\text{im}[\hat{Z}]}=\int_{-\infty}^\infty{G(v)\text{sech}(u-v)dv}\equiv{\int_{-\infty}^{\infty}{G(v)K(u-v)dv}}
\end{equation}
Here $K$ is the kernel, which describes the underlying model. This leads to a deconvolution of the form:
\begin{equation}\label{eq:DRT deconvolution}
    \mathcal{F}(F)=\frac{\mathcal{F}(\hat{I})}{\mathcal{F}(K)}
\end{equation}
Similar methods based on the application of convolution theorem on Fourier, Laplace, and Mellin transforms have been used extensively in the literature on inverse problems in statistical mechanics \cite{Bazant01, Chen90, Hughes90, Lakhtakia84, Weiss59}.

In practice, measurement yields a vector of impedance values ($\textbf{Z}$), which is affected by measurement error. The measured impedance cannot be directly substituted into Equation (\ref{eq:DRT deconvolution}). If the measurement error is a white noise, the Fourier transform of $\textbf{I}=-2\,\text{im}[\textbf{Z}]$ yields a finite value at high frequencies:
\begin{equation}
    \lim_{s\xrightarrow{}{\infty}}\mathcal{F}(\textbf{I})\xrightarrow{}{\text{constant}}
\end{equation}
The Fourier transform of the relaxation kernel vanishes at high frequencies:
\begin{equation}
    \lim_{s\xrightarrow{}{\infty}}\mathcal{F}(K)\xrightarrow{}{0}
\end{equation}
As a consequence, if the measured impedance is substituted directly into Equation (\ref{eq:DRT deconvolution}), the Fourier transform of the DRT diverges:
\begin{equation}
    \lim_{s\xrightarrow{}{\infty}}\mathcal{F}(F)=\lim_{s\xrightarrow{}{\infty}}\frac{\mathcal{F}(\textbf{I})}{\mathcal{F}(K)}\xrightarrow{}{\infty}
\end{equation}
The DRT oscillates wildly, indicating that the problem is ill-posed. The modification proposed by Boukamp concerns the method for suppressing the high-frequency amplitudes induced by the direct substitution of $\mathbf{I}$ into Equation (\ref{eq:DRT deconvolution}):
\begin{equation}
    \mathcal{F}(F)=W(s)\frac{\mathcal{F}(\mathbf{I})}{\mathcal{F}(K)}
\end{equation}
Here $W(s)$ is the window function, which Boukamp defines as:
\begin{equation}
     W(s;\alpha,\beta)=\frac{1}{4}(\tanh[\beta(\alpha+s)]+1)(\tanh[\beta(\alpha-s)]+1)
\end{equation}
The parameters $\alpha$ and $\beta$ describe the width and decay of the window function, respectively. The window function converges to $0$ more rapidly than $\mathcal{F}(K)$, thus restoring the property:
\begin{equation}\label{eq:Vanishing Frequency}
    \lim_{s\xrightarrow{}{\infty}}\mathcal{F}(F)=\lim_{s\xrightarrow{}{\infty}}W(s)\frac{\mathcal{F}(\mathbf{I})}{\mathcal{F}(K)}=0
\end{equation}
We have thus far treated the vector $\textbf{I}$ as if it were a continuous function on $\mathbb{R}$, and thus subject to Fourier transform. In practice, the data set can be made continuous on $\mathbb{R}$ via interpolation and extrapolation, as was done by Boukamp. This procedure is valid, but difficult to explain within the limited scope of the present work. For brevity, the present work employs a closely related formulation based on the discrete Fourier (DF) transform. The DF transform operator ($\mathcal{D}[\cdot]$) acts on a vector of equispaced data as follows:
\begin{equation}
    \left[\mathcal{D}(\textbf{I})\right]_j\equiv\sum_{j'=1}^J\left[\textbf{I}\right]_{j'}\exp\left[-\frac{2\pi{i}(j-1)(j'-1)}{J}\right]
\end{equation}
We define the arrangement operator ($\mathcal{A}$) as:
\begin{equation}\label{eq:Arrangement Operator}
    \left[\mathcal{A}\mathcal{D}(\textbf{I})\right]_{j'}\equiv\begin{cases}
    \left[\mathcal{D}(\textbf{I})\right]_{j-1}\enskip\text{if}\enskip{j'\in\left[0,\frac{J}{2}\right]}\\
    \left[\mathcal{D}(\textbf{I})\right]_{j-1-J}\enskip\text{if}\enskip{j'\in\left[-\frac{J}{2},0\right)}
    \end{cases}
\end{equation}
Note that the indexes $j$ and $j'$ in Equation (\ref{eq:Arrangement Operator}) are defined on $\{1, ...J\}$ and $\{-\left\lfloor{\frac{J}{2}}\right\rfloor, ...\left\lfloor{\frac{J}{2}}\right\rfloor\}$, respectively. The arrangement operator allows direct substitution of $\mathcal{F}[\textbf{I}]$ with $\mathcal{A}\mathcal{D}[\textbf{I}]$.

The Fourier transform algorithm possesses a representation of the form:
\begin{equation}
    G(v;\textbf{a},\textbf{b})\equiv\sum_{m=1}^M{a_m\cos\left(2\pi[m-1]\frac{v}{v_M-v_1}\right)+b_m\sin\left(2\pi[m-1]\frac{v}{v_M-v_1}\right)}
\end{equation}
The parameters $\textbf{a}$ and $\textbf{b}$ are related to the DF transform of the distribution via:
\begin{align}
    \textbf{a}&=\text{re}[\mathcal{D}(\textbf{F})]\\
    \textbf{b}&=\text{im}[\mathcal{D}(\textbf{F})]
\end{align}
Here $\textbf{F}$ is the vector of DRT evaluated at $\textbf{v}$, with $G(v_m)$ as the $m$-th component of $\textbf{F}$. The interior problem is a deconvolution of the form
\begin{equation}\label{eq:Interior Problem Boukamp}
    \textbf{F}^*=\mathcal{D}^{-1}\mathcal{A}^{-1}\left[W(s;\alpha,\beta)\mathcal{A}\mathcal{D}(\textbf{I})\oslash\mathbb{K}(\textbf{s})\right]
\end{equation}
The symbol $\oslash$ is an element-wise division with $\mathbb{K}\equiv\mathcal{F}(K)$ evaluated at $\mathbf{s}=\frac{\textbf{j'}}{v_M-v_1}$, where $j'$ is the index of $\mathcal{A}\mathcal{D}(\textbf{I})$, as defined in Equation (\ref{eq:Arrangement Operator}). Equation (\ref{eq:Interior Problem Boukamp}) implies that $J=M$, since the size of $\textbf{F}^*$ must be equal to $\textbf{I}$. Note that for $\textbf{F}^*$ to occur in order of increasing $t$, the data set $\mathbf{I}$ needs to be input in order of increasing $u$.

The difficulty faced by Boukamp concerns the selection of window parameters $\alpha$ and $1/\beta$, which we identify as the complexity control parameters of the DRT inversion algorithm. This is sensible, given that increasing $\alpha$ and decreasing $\beta$ increases the amplitude of high-frequency terms included in $\textbf{F}^*$. The objective, then, is to construct an exterior problem to determine appropriate values of $\alpha$ and $\beta$.

This can be neatly achieved by utilizing the real part of the data set in a validation step:
\begin{equation}
    \alpha^*,\beta^*\equiv\underset{\alpha,\beta}{\text{argmin}}\enskip\left\lVert{\text{re}\left[\textbf{Z}-\hat{\textbf{Z}}(\textbf{F}^*)\right]}\right\rVert_2^2
\end{equation}
This exterior problem requires the real part of the measurement error to be independent of the imaginary part of the measurement error.

We demonstrate the effectiveness of the proposed exterior problem using the sample problem of Schichlein, M\"{u}ller, Voigts, Kr\"{u}gel, and Ivers-Tiff\'{e}e \cite{Schichlein02}, which consists of three Debye and three Cole-Cole elements connected in series:
\begin{equation}
    \hat{Z}(u;\textbf{R},\boldsymbol{\mu},\boldsymbol{\alpha})=\sum_{m=1}^6{\frac{R_m}{1+\left[i\exp(\mu_m-u)\right]^{\alpha_m}}}
\end{equation}
The parameter values ($\textbf{R},\boldsymbol{\mu},\boldsymbol{\alpha}$) of the sample problem are reported in Table \ref{tab:Sample Problem Boukamp}. The underlying distribution takes the form:
\begin{equation}
    G(v)=\sum_{m=1}^3{R_m\delta(v-\mu_m)}+\sum_{m=4}^6{\frac{R_m}{2\pi}\frac{\sin[(1-\alpha_m)\pi]}{\cosh[\alpha_m(v-\mu_m)]}-\cos[(1-\alpha_m)\pi]}
\end{equation}
We generate synthetic data for angular frequencies in the domain $\omega={10}^{-2}$ to $10^{6}$ rad/s with 10 data points per decade.

\begin{table}[h!]
    \centering
    \begin{tabular}{|c|c|c|c|c|c|c|}
    \hline
    $m$ & 1 & 2 & 3 & 4 & 5 & 6\\
    \hline
    $R_m$ & 1 & 0.5 & 1 & 0.5 & 1 & 1\\
    $\mu_m$ & -11.5 & -9.2 & -6.9 & -4.6 & -2.3 & 0\\
    $\alpha_m$ & 1 & 1 & 1 & 0.85 & 0.8 & 0.75\\
    \hline
    \end{tabular}
    \caption{List parameter values used for the sample problem. The first three element, i.e., $m=1, ...3$ have $\alpha_m=1$, and are thus Debye elements.}
    \label{tab:Sample Problem Boukamp}
\end{table}

\begin{figure*}[h!]
    \centering
    \includegraphics[scale=0.6]{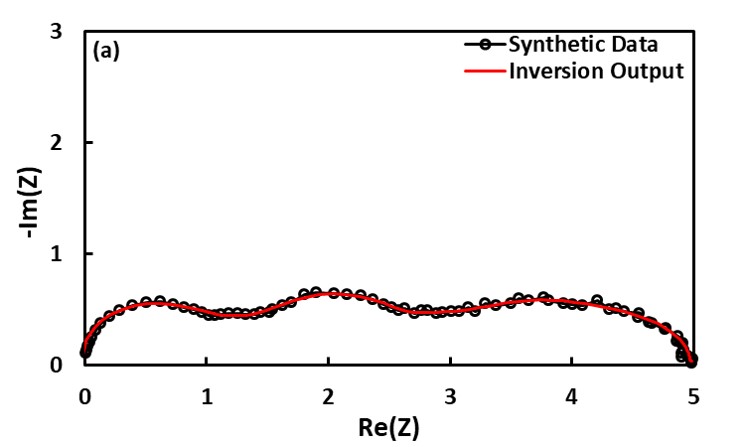}
    \includegraphics[scale=0.6]{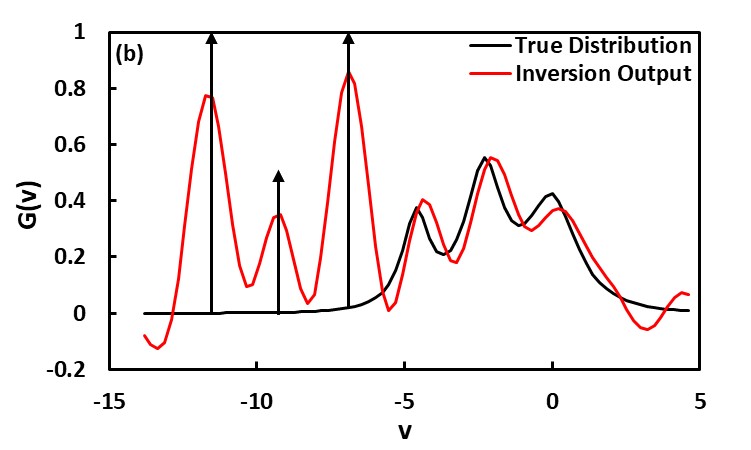}
    \caption{(a) Nyquist plot of the impedance data as well as the prediction obtained from the inversion output of the Fourier transform algorithm. The noise takes the form $Z=\hat{Z}+|\hat{Z}|\times{0.005}\times(N(0,1)+iN(0,1))$. (b) True and inverted distribution obtained using proposed exterior problem. The black arrows indicate the location and magnitude of the Dirac delta distributions.}
    \label{fig:Sample Problem Boukamp}
\end{figure*}

The inversion result is reported in Figure \ref{fig:Sample Problem Boukamp}. The Fourier transform algorithm is clearly able to distinguish the six characteristic timescales of the process, as well as the distribution masses associated with each characteristic timescale. However, the algorithm overestimates the variances associated with the Debye elements. This is because it is impossible to reproduce the high-frequency component of the Dirac delta distributions while also preventing the interior problem from becoming ill-posed. Accurate reproduction of the Dirac delta distributions require the Cole-Cole distributions to be highly oscillatory; accurate reproduction of the Cole-Cole distributions require the Dirac delta distributions to be excessively smoothed. This concept is critical to the design of EIS inversion algorithms, and will be expanded upon in Section \ref{sec:Design of Algorithm}. 

\subsection{L-Curve Algorithm}\label{subsec:L-Curve Algorithm}
The L-curve algorithm employs fixed-mesh triangular basis function representation, i.e., Equation (\ref{eq:Triangular Basis Representation}). The interior problem is the minimization of the regularized misfit:
\begin{equation}\label{eq:Interior Problem Florsch}
    \textbf{g}^*\equiv\underset{\textbf{g}}{\text{argmin}}\enskip\underbrace{\left\lVert{\textbf{Z}-\hat{\textbf{Z}}}\right\rVert_2^2}_{\text{misfit}}+\underbrace{\lambda\underbrace{\left\lVert{\textbf{L}_1\textbf{g}}\right\rVert_2^2}_{\text{complexity}}}_{\text{penalty}}\equiv\underset{\textbf{g}}{\text{argmin}}\enskip\mathcal{M}(\mathbf{g})+\lambda\mathcal{P}(\mathbf{g})
\end{equation}
The complexity control parameter is the inverse of the regularization parameter ($1/\lambda$), and the complexity is the first derivative of the distribution ($\left\lVert{\textbf{L}_1\textbf{g}}\right\rVert_2^2$). 

\begin{figure}[h!]
    \centering
    \includegraphics[scale=0.42]{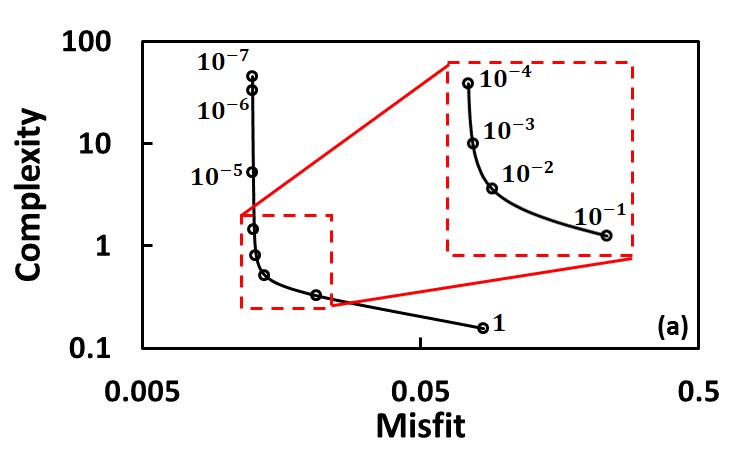}
    \includegraphics[scale=0.42]{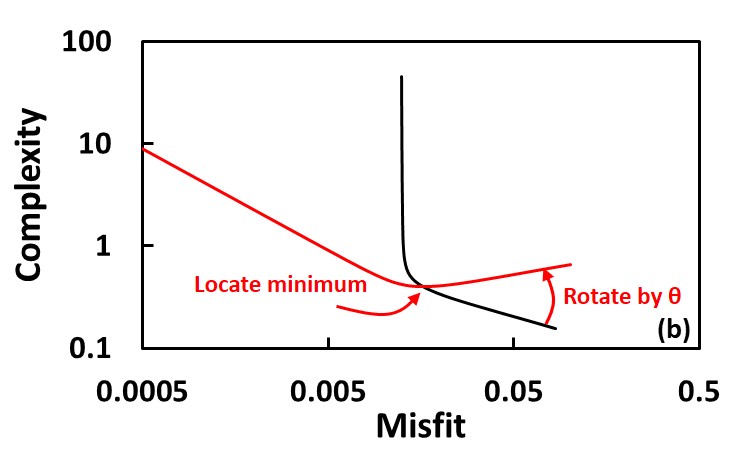}
    \includegraphics[scale=0.42]{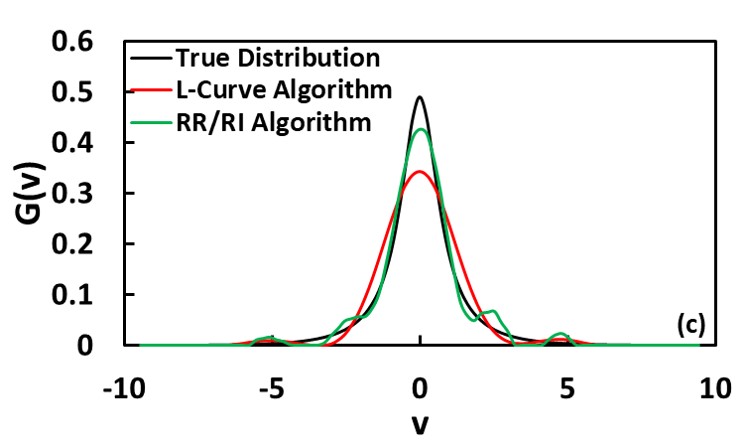}
    \caption{(a) The L-curve corresponding to the unimodal Cole-Cole distribution reported in Figure \ref{fig:Parameter Reduction} (a). The corner is identifiable but ambiguous. (b) Rotation of the L-curve yields an unambiguous minimum. (c) True and inverted distribution obtained using the proposed exterior problem. The inversion output of the RR/RI algorithm is included as a reference.}
    \label{fig:L-Curve Illustration}
\end{figure}

The conventional definition of the L-curve, which may be attributed to Hansen \cite{Hansen93}, is a log-log plot of the complexity against the misfit, parameterized by the regularization parameter. For illustration, we have generated the L-curve corresponding to the unimodal Cole-Cole distribution reported in Figure \ref{fig:Parameter Reduction} (a), which is shown in Figure \ref{fig:L-Curve Illustration} (a). The L-curve algorithm claims that the appropriate value of the regularization parameter is that which corresponds to the corner of the L-curve, i.e., somewhere in between 0.1 and 0.0001. This range compares favorably with the optimal regularization parameter of $7.1\times10^{-4}$ obtained using the RR/RI algorithm, but suffers from ambiguity in the definition of a corner. This issue is partially resolved by Florsch, Revil, and Camerlynck by defining the corner of an L-curve as the minimum obtained when the L-curve is rotated clockwise, as illustrated in Figure \ref{fig:L-Curve Illustration} (b). The exterior problem takes the form:
\begin{equation}\label{eq:L-Curve Rotation}
    \lambda^*\equiv\underset{\lambda}{\text{argmin}}\enskip\cos(\theta)\,\ln\left[\mathcal{M}(\textbf{g}^*)\right]-\sin(\theta)\,\ln\left[\mathcal{P}(\textbf{g}^*)\right]
\end{equation}
This exterior problem is incomplete, since the angle of rotation ($\theta$) remains unknown. 

The determination of $\theta$ is challenging due to the lack of a statistically meaningful interpretation for the rotation of the L-curve. Our objective is to manipulate the exterior problem into a form which carries statistical sense, and thus determine an appropriate value for $\theta$. Applying the first-order optimality condition to the objective function of Equation (\ref{eq:Interior Problem Florsch}),
\begin{equation}
    \lambda=-\frac{d\mathcal{M}}{d\mathcal{P}}(\mathbf{g}^*)
\end{equation}
This establishes $\textbf{g}^*$ as a function of $\lambda$. This relationship must, in particular, hold true when $\lambda=\lambda^*$:
\begin{equation}
    \lambda^*=-\frac{d\mathcal{M}}{d\mathcal{P}}(\lambda^*)
\end{equation}
Applying the first-order optimality condition to the objective function of Equation (\ref{eq:L-Curve Rotation}),
\begin{equation}
    \cot{\theta}=-\frac{\mathcal{P}(\lambda^*)}{\mathcal{M}(\lambda^*)}\frac{d\mathcal{M}}{d\mathcal{P}}(\lambda^*)
\end{equation}
Combining the previous two expressions,
\begin{equation}
    \cot{\theta}=\frac{\lambda^*\mathcal{P}(\lambda^*)}{\mathcal{M}(\lambda^*)}
\end{equation}
This establishes a non-obvious relationship between $\theta$ and the relative size of the misfit and the penalty. By requiring the misfit to be of the same order as the penalty, as argued by Hansen \cite{Hansen99}, we obtain the condition $\theta^*=\pi/4$. Application of this angle of rotation to the L-curve of Figure \ref{fig:L-Curve Illustration} (a) yields $\lambda^*=0.040$, which corresponds to the green curve shown in Figure \ref{fig:L-Curve Illustration} (c). This choice of regularization parameter results in a DRT which is less oscillatory than that of the RR/RI algorithm, but also relatively over-smoothed.

Florsch, Revil, and Camerlynck have defined the L-curve as the log-log plot of $\int_{-\infty}^{\infty}G(v)dv$ against the misfit, as opposed to the conventional pairing of complexity and misfit. The analytical justification for the efficacy of the L-curve requires the conventional pairing to be used \cite{Hansen99}. The use of an unconventional definition accounts for the distinct lack of L shape in the curves reported by Florsch, Revil, and Camerlynck, and the attendant difficulty in determining $\lambda^*$. The supplemented exterior problem introduced in the present work is intended for the conventional pairing. We note that a well-established exterior problem based on the identification of the point of maximum curvature exists \cite{Hansen93}, and is also intended for the conventional pairing. 

\subsection{Maximum Entropy Algorithm}\label{subsec:Maximum Entropy Algorithm}
H\"{o}rlin applied the maximum entropy algorithm to a modified DRT problem obtained by substituting $E(v)\equiv{G(v)\exp(-v/2)}$ and $\hat{\xi}(u)\equiv{\hat{Z}(u)/\sqrt{\omega}}$ into Equation (\ref{eq:DRT convolution}):
\begin{equation}
    \hat{\xi}(u;E(v))=\int_{-\infty}^{\infty}E(v)\frac{\exp\left(-\frac{1}{2}[v-u]\right)-i\exp\left(\frac{1}{2}[v-u]\right)}{2\cosh(v-u)}dv
\end{equation}
This does not change the essential nature of the problem, but modifies the appearance of the DRT in a useful way. According to this integral, the Warburg impedance corresponds to a uniform distribution:
\begin{equation}\label{eq:DRT of Planar Warburg}
    \hat{Z}=\frac{R}{\sqrt{i\omega}}\Longleftrightarrow{E(v)=\frac{R}{\pi}}
\end{equation}
The Debye relaxation process corresponds to a Dirac delta distribution:
\begin{equation}
    \hat{Z}=\frac{R}{1+i\omega\exp(\mu_0)}\Longleftrightarrow{E(v)=R\exp\left(\frac{\mu_0}{2}\right)\delta(v-\mu_0)}
\end{equation}
Here $\mu_0\equiv\ln(\tau_0)$ is the log-characteristic timescale of the relaxation process. Warburg and relaxation processes are more discernible on $E(v)$ than $G(v)$.

The maximum entropy algorithm employs a fixed-mesh triangular basis function representation:
\begin{equation}
    E(v)\equiv{\sum_{m=1}^M{e_m\Delta_m(v)}}
\end{equation}
The interior problem takes the form:
\begin{align}
    \textbf{e}^*&\equiv\underset{\textbf{e}}{\text{argmin}}\enskip{\mathcal{U}(\textbf{e})-T\,\mathcal{S}(\textbf{e})}\\
    \mathcal{U}(\textbf{e})&\equiv\frac{\left\lVert{\textbf{Z}-\hat{\textbf{Z}}(\textbf{e})}\right\rVert_2^2}{2J\sigma^2}\\
    \label{eq:Entropy Definition}
    \mathcal{S}(\textbf{e})&\equiv\text{sum}\left[\textbf{e}-\textbf{e}\circ\ln(\textbf{e})\right]
\end{align}
Here $\mathcal{U}$ is the misfit, $T$ is the temperature, $\mathcal{S}$ is the entropy, $\sigma^2$ is the measurement variance, and $\text{sum}[\cdot]$ is the vector summation operator. We use the symbols $\mathcal{U}$, $T$, and $\mathcal{S}$ to emphasize the connection between the algorithm and thermodynamics. The misfit $\mathcal{U}$ is the internal energy of the system, which is composed of quadratic interactions between the experimental data and the model prediction. The system tends to collapse to the state with the lowest internal energy, which corresponds to an ill-posed problem, but is prevented from doing so by the temperature, which weights the contribution of entropy. For a fixed distribution mass, the entropy is lowest when the distribution is a Dirac delta, and highest when the distribution is uniform. 

The complexity control parameter is the inverse temperature ($1/T$) and the complexity is entropy ($\mathcal{S}$). With increasing temperature, the contribution of entropy increases, thus yielding broader distributions which may be thought of as simpler. The exterior problem is a constraint on the misfit:
\begin{equation}\label{eq:Exterior Problem Horlin}
    \mathcal{U}(\textbf{e})=1
\end{equation}
This exterior problem is used to calculate the optimal temperature ($T^*$). The evaluation of the misfit requires the measurement variance to be known, which may be obtained from repeat measurements, or by comparison with the Kramers-Kronig transform of the impedance spectra. 

\begin{figure*}[h!]
    \centering
    \includegraphics[scale=0.42]{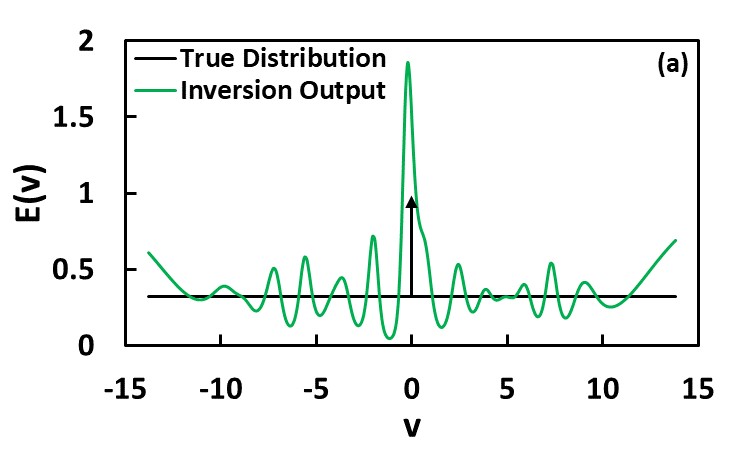}
    \includegraphics[scale=0.42]{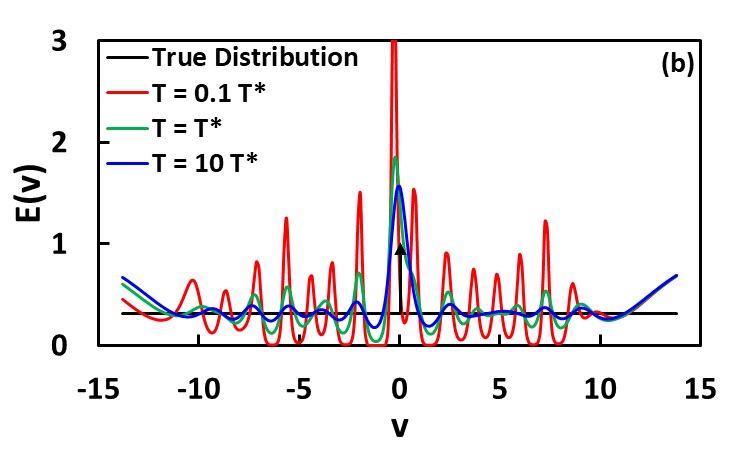}
    \includegraphics[scale=0.42]{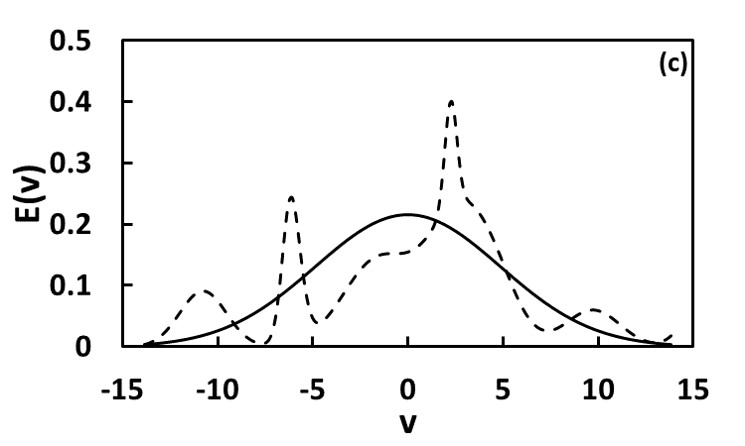}
    \caption{(a) True and inverted distribution obtained using the maximum entropy algorithm.  The noise takes the form $Z=\hat{Z}+|\hat{Z}|\times{0.005}\times(N(0,1)+iN(0,1))$. (b) Effect of temperature on the inverted distribution. (c) We intuitively perceive the dotted multimodal distribution to be more complicated than the solid unimodal distribution. However, the entropy of the two distributions are equal. }
    \label{fig:Maximum Entropy Illustration}
\end{figure*}

We illustrate the difficulty faced by the maximum entropy algorithm through an example, wherein the the true impedance takes the form:
\begin{equation}
    \hat{Z}(\omega)=\frac{1}{1+i\omega}+\frac{1}{\sqrt{i\omega}} \Longleftrightarrow{ E(v)=\delta(v)+\frac{2}{\pi}}
\end{equation}
We collect 10 log-equispaced synthetic data per decade over the angular frequency range $\omega_1={10}^{-4}\enskip\text{rad/s}$ and $\omega_J={10}^4\enskip\text{rad/s}$. The resulting probability distribution is inverted over the domain $\tau_1={10}^{-6}$ and $\tau_M={10}^6$, as shown in Figure \ref{fig:Maximum Entropy Illustration} (a). The uniform distribution associated with the Warburg element has dispersed into a large number of peaks which are difficult to distinguish from Debye elements with relatively small resistances. As illustrated in Figure \ref{fig:Maximum Entropy Illustration} (b), modifying the temperature does not yield a strict improvement in the result. At lower temperatures, the Dirac delta distribution becomes more prominent, but this occurs at the expense of increasing the overall oscillation of $E(v)$. At higher temperatures, the algorithm reproduces the uniform part of $E(v)$ with greater accuracy, but the Dirac delta distribution is suppressed. 

The simultaneous reproduction of simple and complex features in a distribution is the central challenge in the design of EIS inversion algorithms. This problem is not unique to the maximum entropy algorithm, as evidenced by the inability of the Fourier transform algorithm to simultaneously reproduce the Dirac delta and the Cole-Cole distributions (see Section \ref{subsec:Fourier Transform Algorithm}), and has been repeatedly observed in the literature. For example, application of the RR/RI algorithm on the fractal element, whose DRT is characterized by a smooth domain terminated by a discontinuity, has been reported to output an oscillatory distribution \cite{Saccoccio14}. Likewise, the adaptive multi-parameter regularization approach of \v{Z}ic, Pereverzyev, Suboti\'{c}, and Pereverzyev returns an oscillatory distribution for the fractal element \cite{Zic20}. Dion and Lasia are able to invert the fractal element with minimal oscillation using a generic inversion toolbox \cite{Dion99}. However, the interior and exterior problems used are not reported. To our knowledge, no DRT inversion algorithm has addressed the simultaneous reproduction of simple and complex features comprehensively, although the ISGP algorithm \cite{Hershkovitz11a, Hershkovitz11b} and the hierarchical Bayesian algorithm \cite{Ciucci15} have provided partial resolution to this problem. 

The difficulty in simultaneously reproducing the uniform and Dirac delta distributions using the maximum entropy algorithm may be fundamentally attributed to the use of entropy as complexity, which aligns with our intuition, but only in the limit as $T\xrightarrow{}\infty$, which yields a uniform distribution. To illustrate this point, consider the two distributions shown in Figure \ref{fig:Maximum Entropy Illustration} (c), which possess equal entropy. We intuitively perceive the dotted multimodal distribution to be more complicated than the solid unimodal distribution, in contradiction to the value of entropy assigned by Equation (\ref{eq:Entropy Definition}). As a consequence, the distribution reported in Figure \ref{fig:Maximum Entropy Illustration} (a) may be simple from an entropic point-of-view, but complex according to intuition. Rectification of the maximum entropy algorithm would likely require entropy to be re-defined.

\section{Design of Algorithm}\label{sec:Design of Algorithm}
Thus far, complexity has been described in terms of intuition, and has been associated with the idea of smoothness in distributions. This is true of most of the algorithms listed in Section \ref{sec:Sample Algorithms} and Section \ref{sec:Analysis}. The RR/RI algorithm of Section \ref{subsec:RR/RI Algorithm} considers the first derivative of the distribution as the complexity, and thus smooth distributions with low first derivatives are simple. The Fourier transform algorithm of Section \ref{subsec:Fourier Transform Algorithm} considers the high-frequency terms of $G(v)$ as the complexity, whose suppression leads to simple, smooth distributions. In this section, we argue that the intuitive understanding of complexity is incorrect, and that it should be described in terms of the physics implied by the distribution, and guided by Occam's razor.

\begin{figure*}[h!]
    \centering
    \includegraphics[scale=0.6]{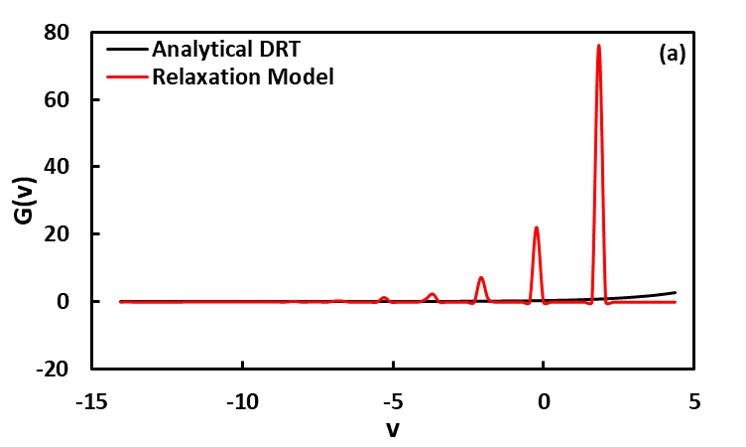}
    \includegraphics[scale=0.6]{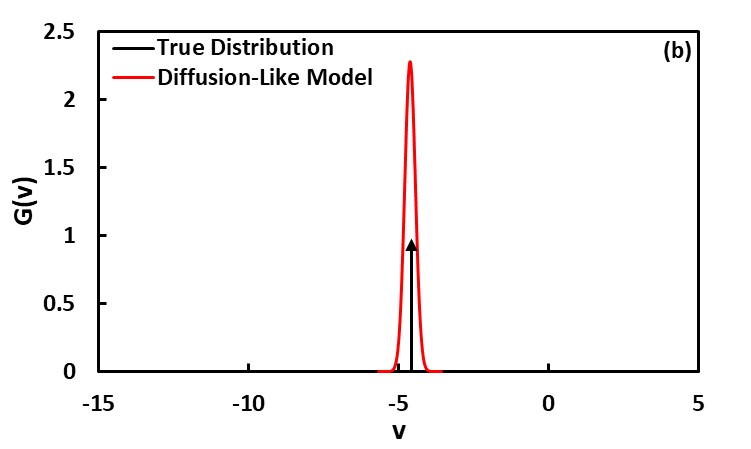}
    \caption{(a) The DRT and (b) the DDT of a planar Warburg element. The DRT is obtained using the RR/RI algorithm, while the DDT is obtained using the gEISi algorithm introduced in Section \ref{subsec:gEISi Algorithm} set to a planar Warburg model. For completeness, we include the analytical DRT of the planar Warburg element, i.e., Equation (\ref{eq:DRT of Planar Warburg}). The solution for $v\geq{4.34}$ has been excluded to improve visibility. The true impedance corresponds to a planar Warburg model with $G(v)=\delta(v-\ln(0.01)).$}
    \label{fig:Model Simplicity}
\end{figure*}

The works of Florsch, Revil, and Camerlynck \cite{Florsch14} and Song and Bazant \cite{Song18} suggests that when an appropriate underlying model is used, the inverted distribution tends to appear less complicated. Figure \ref{fig:Model Simplicity} illustrates (a) the DRT and (b) the DDT obtained when a Warburg element is inverted using a relaxation and a diffusion-like model, respectively. The diffusion-like model returns a less complicated distribution relative to the relaxation model, suggesting the relative propriety of the diffusion-like model. 

\begin{figure*}
    \centering
    \includegraphics[scale=0.42]{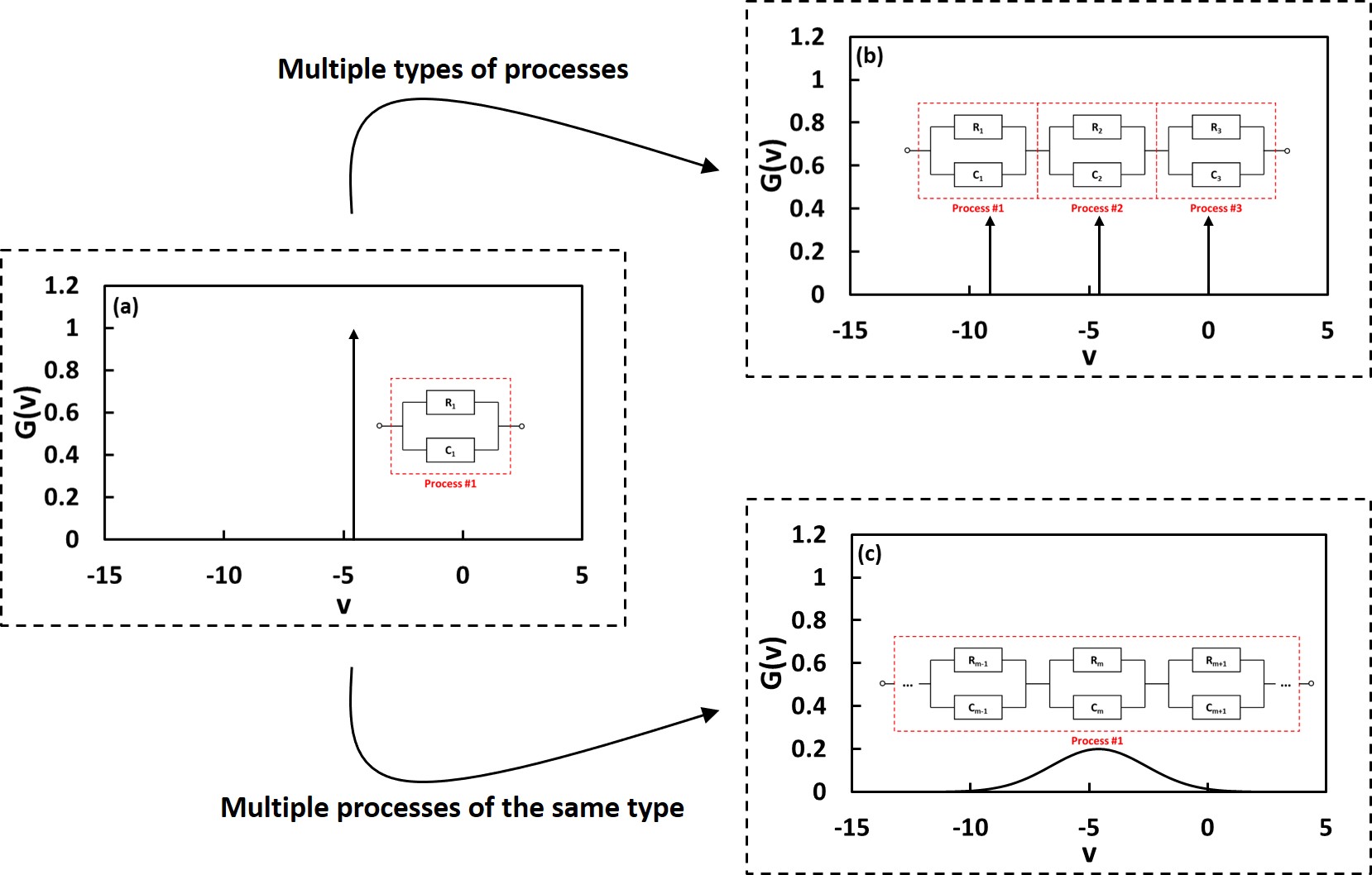}
    \caption{A phenomenological picture of complexity in EIS inversion problems. (a) The Dirac delta distribution is the simplest possible distribution. (b) Increase in modality and (c) variance correspond to an increase in complexity.}
    \label{fig:Physical Interpretation}
\end{figure*}

We draw this observation to its logical conclusion, wherein the application of an exactly correct underlying model results in the simplest possible distribution. That is, if the impedance arises due to an integral over the underlying model:
\begin{equation}
    \hat{Z}(\omega)=\int_{-\infty}^{\infty}G(v)\hat{z}(\omega,v)dv
\end{equation}
then the simplest distribution is obtained when $\hat{Z}(\omega)=\hat{z}(\omega,\mu_0)$, for which $G(v)=\delta(v-\mu_0)$. This leads to the counterintuitive conclusion that \textit{the simplest possible distribution is the Dirac delta distribution}. This conclusion is consistent with Occam's razor; as illustrated in Figure \ref{fig:Physical Interpretation}, for an underlying relaxation model, the Dirac delta distribution corresponds to a single relaxation process. Any other distribution would imply various combinations of multiple types of relaxation processes (Figure \ref{fig:Physical Interpretation} (b)) and multiple relaxation processes of the same type (Figure \ref{fig:Physical Interpretation} (c)). As a corollary, \textit{complexity increases with modality and variance}.

The idea of the Dirac delta distribution as the zero point of complexity in EIS inversion problems contradicts much of the EIS inversion literature. The Fourier transform algorithm \cite{Boukamp15, Schichlein02}, L-curve algorithm \cite{Florsch14}, maximum entropy algorithm \cite{Horlin98}, the RR/RI algorithm \cite{Saccoccio14, Han15}, the hierarchical Bayesian algorithm \cite{Ciucci15}, as well as several other unmentioned algorithms \cite{Song18, Tuncer01} all perceive the Dirac delta distribution as possessing maximum complexity. As a consequence, these algorithms are not able to identify the Dirac delta distribution when it occurs in conjunction with smooth features, as illustrated in Section \ref{subsec:Fourier Transform Algorithm} and Section \ref{subsec:Maximum Entropy Algorithm}. 

In the remainder of Section \ref{sec:Design of Algorithm}, we will use this phenomenological picture of complexity to construct an improved EIS inversion algorithm, applicable to the generalized EIS inversion problem, subject to constraints on the allowed computational time. We will also consider the construction of credible intervals for the parameters and distributions arising from the algorithm. We apply the algorithm to a DRT problem exhibiting a wide range of smoothness, and compare its performance with the algorithms discussed in Section \ref{sec:Sample Algorithms}. We also apply the algorithm to three generalized EIS inversion problems taking the form of a distributed Randles circuit \cite{Randles47} possessing nearly overlapping distribution peaks.

\subsection{Generalized EIS Inversion Problem}\label{subsec:Generalized EIS Inversion Problem}
The generalized EIS inversion problem is an abstraction of the DRT problem, in which the underlying model is specified, but not assumed a priori:
\begin{equation}\label{eq:generalized EIS Problem}
    \hat{Z}=\hat{Z}(\omega;\boldsymbol{\beta},\{G_l(v)\})
\end{equation}
The model $\hat{Z}$ is parameterized by $\boldsymbol{\beta}$, which takes on definite values, and $\{G_l(v)\}$, which is a collection of distributions. Our objective is to construct a generalized EIS inversion algorithm which estimates the parameters $\boldsymbol{\beta}$ and $\{G_l\}$ given any such model. The generalized EIS inversion problem includes models taking the form of a Fredholm integral of the second kind, which necessitates an iterative solver. 

To illustrate the generalized EIS inversion problem, consider the Randles circuit \cite{Randles47}:
\begin{align}
    \hat{Z}(\omega)&=\frac{R_{CT}}{\frac{R_{CT}}{R_{CT}+\hat{Z}_D}+i\omega\tau_{CT}}+R_\infty\\
    \hat{Z}_D(\omega)&=R_D\frac{\tanh{\sqrt{i\omega\tau_D}}}{\sqrt{i\omega\tau_D}}
\end{align}
The Randles circuit describes a charge-transfer process with resistance $R_{CT}$ and timescale $\tau_{CT}$ occurring in series with a diffusion process with resistance $R_D$ and timescale $\tau_D$, both located within an electrical double layer. Due to surface inhomogeneity, $\tau_D$ is distributed, leading to a diffusion impedance of the form:
\begin{equation}\label{eq:Distributed Diffusion}
    \hat{Z}_D^{-1}(\omega;G_D(v))=\int_{-\infty}^{\infty}G_D(v)\frac{\sqrt{i\omega\exp(v)}}{\tanh{\sqrt{i\omega\exp(v)}}}dv
\end{equation}
The remainder of the model is recast in the form of the DRT problem:
\begin{equation}\label{eq:Distributed Charge Transfer}
    \hat{Z}(\omega;R_\infty,G_{CT}(v),G_D(v))=\int_{-\infty}^{\infty}\frac{G_{CT}(v)}{\frac{R_{CT}}{R_{CT}+\hat{Z}_D(\omega;G_D(v))}+i\omega\exp(v)}dv+R_\infty
\end{equation}
Equation (\ref{eq:Distributed Diffusion}) and Equation (\ref{eq:Distributed Charge Transfer}) constitute a single model of the generalized form, with $R_\infty$ as $\boldsymbol{\beta}$, and $G_{CT}(v)$ and $G_D(v)$ as $\{G_l(v)\}$. The generalized EIS inversion problem corresponding to the Randles circuit contains an embedded integral, which is incompatible with linearly constrained quadratic programming. Furthermore, it is a Fredholm integral of the second kind.

\subsection{Idealized EIS Inversion Algorithm}\label{subsec:Idealized EIS Inversion Algorithm}
The idealized form of the generalized EIS inversion algorithm employs a floating-mesh Gaussian basis function representation:
\begin{equation}
    G_l(v;\{\textbf{R}_l,\boldsymbol{\mu}_l,\boldsymbol{\nu}_l\})\equiv\sum_{m_l=1}^{M_l}\frac{R_{m_l}}{\sqrt{2\pi}\exp(\nu_{m_l}/2)}\exp\left[-\frac{(v-\mu_{m_l})^2}{2\exp(\nu_{m_l})}\right]
\end{equation}
Here $R_{m_l}$, $\mu_{m_l}$, and $\nu_{m_l}$ are the mass, mean, and log-variance of the $m_l^\text{th}$ basis function of the $l^\text{th}$ distribution. Typically, the use of a floating-mesh representation increases the computational time required to solve the problem, since it is incompatible with linearly constrained quadratic programming. However, as illustrated by the Randles circuit, the generalized EIS inversion problem will be largely incompatible with linearly constrained quadratic programming to begin with, and so this increase in computational time may be thought of as a sunk cost. In fact, we expect the reduction in the number of parameters effected by the use of the floating-mesh representation to reduce the overall computational time. The use of Gaussian basis function follows the observation made by Han, Saccoccio, Chen, and Ciucci \cite{Han15} concerning the effectiveness of radial basis functions.

We adapt the interior problem from the RR/RI algorithm of Saccoccio, Han, Chen, and Ciucci \cite{Saccoccio14}. The real part of the idealized interior problem takes the form:
\begin{align}\label{eq:Real Idealized Interior Problem Effendy}
    \boldsymbol{\beta}^\mathbb{R},\{\textbf{R}_l,\boldsymbol{\mu}_l,\boldsymbol{\nu}_l\}^\mathbb{R},\nu_\epsilon^\mathbb{R}&\equiv\underset{\boldsymbol{\beta},\{\textbf{R}_l,\boldsymbol{\mu}_l,\boldsymbol{\nu}_l\},\nu_\epsilon}{\text{argmin}}\enskip\mathcal{M}^\mathbb{R}+\mathcal{P}^\mathbb{R}+\mathcal{H}^\mathbb{R}\\
    \mathcal{M}^\mathbb{R}&\equiv\frac{\left\lVert{\text{re}[\textbf{Z}-\hat{\textbf{Z}}]\,\oslash\,\text{abs}[\textbf{Z}]}\right\rVert_2^2}{\exp(\nu_\epsilon)}+\frac{\text{sum}\left[\text{im}[\textbf{Z}-\hat{\textbf{Z}}]\,\oslash\,\text{abs}[\textbf{Z}]\right]^2}{J\exp(\nu_\epsilon)}\\
    \mathcal{P}^\mathbb{R}&\equiv\lambda_\mathcal{N}\sum_{l=1}^L\mathcal{N}[G_l(v)]+\lambda_\nu\sum_{l=1}^L\left\lVert{\exp(\boldsymbol{\nu}_l)}\right\rVert_1\\
    \mathcal{H}^\mathbb{R}&\equiv\frac{(\mu_\epsilon-\nu_\epsilon)^2}{\sigma_\epsilon^2}+\nu_\epsilon(J+1)
\end{align}
Here $\mathcal{M}$, $\mathcal{P}$, and $\mathcal{H}$ are the misfit, penalty, and hyperprior on the measurement error, $\lambda_\mathcal{N}$ and $\lambda_\nu$ are the regularization parameters on the modality and the variance of the distributions, $\mu_\epsilon$ and $\sigma_\epsilon^2$ are the hyperprior expectation and variance of the log-variance of measurement error, and $\nu_\epsilon$ is the log-variance of measurement error. We calculate the misfit using the real part of the data set and the average of the imaginary part of the data set to account for models containing pure capacitive terms \cite{Florsch14}. The imaginary part of the interior problem takes the symmetric form:
\begin{align}\label{eq:Imaginary Idealized Interior Problem Effendy}
    \boldsymbol{\beta}^\mathbb{I},\{\textbf{R}_l,\boldsymbol{\mu}_l,\boldsymbol{\nu}_l\}^\mathbb{I},\nu_\epsilon^\mathbb{I}&\equiv\underset{\boldsymbol{\beta},\{\textbf{R}_l,\boldsymbol{\mu}_l,\boldsymbol{\nu}_l\},\nu_\epsilon}{\text{argmin}}\enskip\mathcal{M}^\mathbb{I}+\mathcal{P}^\mathbb{I}+\mathcal{H}^\mathbb{I}\\
    \mathcal{M}^\mathbb{I}&\equiv\frac{\left\lVert{\text{im}[\textbf{Z}-\hat{\textbf{Z}}]\,\oslash\,\text{abs}[\textbf{Z}]}\right\rVert_2^2}{\exp(\nu_\epsilon)}+\frac{\text{sum}\left[\text{re}[\textbf{Z}-\hat{\textbf{Z}}]\,\oslash\,\text{abs}[\textbf{Z}]\right]^2}{J\exp(\nu_\epsilon)}
\end{align}
The inverse of the regularization parameters, i.e., $1/\lambda_\mathcal{N}$ and $1/\lambda_\nu$ are the complexity control parameters. The complexity is the modality and the variance of the distributions, following the phenomenological picture of complexity outlined in Figure \ref{fig:Physical Interpretation}. 

The idealized exterior problem is the real-imaginary cross-validation routine:
\begin{equation}\label{eq:Idealized Exterior Problem Effendy}
    \lambda_\mathcal{N}^*,\lambda_\nu^*\equiv\underset{\lambda_\mathcal{N},\lambda_\nu}{\text{argmin}}\enskip\mathcal{M}^\mathbb{R}\left(\boldsymbol{\beta}^\mathbb{I},\{\textbf{R}_l,\boldsymbol{\mu}_l,\boldsymbol{\nu}_l\}^\mathbb{I}\right)+\mathcal{M}^\mathbb{I}\left(\boldsymbol{\beta}^\mathbb{R},\{\textbf{R}_l,\boldsymbol{\mu}_l,\boldsymbol{\nu}_l\}^\mathbb{R}\right)+\mathcal{H}^\mathbb{R}(\nu_\epsilon^\mathbb{I})+\mathcal{H}^\mathbb{I}(\nu_\epsilon^\mathbb{R})
\end{equation}

\subsection{Generalized EIS Inversion Algorithm}\label{subsec:gEISi Algorithm}
An application of the idealized generalized EIS inversion algorithm to an experimental data set modelled by the distributed Randles circuit results in an unacceptably high computational time. The difficulty lies in the discrete nature of the $\mathcal{N}[\cdot]$ operator, which prevents the use of efficient gradient-based methods, and the inability to obtain initial guesses for $\lambda_\mathcal{N}$ and $\lambda_\nu$. We obtain a reasonable compromise between accuracy and computational time by using a proxy measure for modality, and eliminating the variance of the distribution from the complexity.

The simplified interior problem takes the form:
\begin{align}\label{eq:Real Interior Problem Effendy}
     \boldsymbol{\beta}^\mathbb{R},\{\textbf{R}_l,\boldsymbol{\mu}_l,\boldsymbol{\nu}_l\}^\mathbb{R},\nu_\epsilon^\mathbb{R}&\equiv\underset{\boldsymbol{\beta},\{\textbf{R}_l,\boldsymbol{\mu}_l,\boldsymbol{\nu}_l\},\nu_\epsilon}{\text{argmin}}\enskip\mathcal{M}^\mathbb{R}+\mathcal{H}^\mathbb{R}\\\label{eq:Imaginary Interior Problem Effendy}
     \boldsymbol{\beta}^\mathbb{I},\{\textbf{R}_l,\boldsymbol{\mu}_l,\boldsymbol{\nu}_l\}^\mathbb{I},\nu_\epsilon^\mathbb{I}&\equiv\underset{\boldsymbol{\beta},\{\textbf{R}_l,\boldsymbol{\mu}_l,\boldsymbol{\nu}_l\},\nu_\epsilon}{\text{argmin}}\enskip\mathcal{M}^\mathbb{I}+\mathcal{H}^\mathbb{I}
\end{align}
The simplified control parameter is the total number of basis functions used:
\begin{equation}\label{eq:Complexity Effendy}
    M\equiv\sum_{l=1}^L{M_L}
\end{equation}
This control parameter appears to be a reasonable approximation of modality. In the second validation problem of Section \ref{subsec:General Validation}, the Cole-Cole distributions of the charge transfer and diffusion processes are reasonably approximated by a small number of Gaussian basis functions, suggesting that $M$ will be of the same order as the modality for many problems of practical interest. The elimination of variance from the complexity results in a slight widening of the inversion output; this effect is observed in Figure \ref{fig:Model Simplicity} (b), wherein a small but nonvanishing variance is observed, which would have been otherwise absent for the idealized generalized EIS inversion algorithm.

The simplified exterior problem takes the form:
\begin{equation}\label{eq:Exterior Problem Effendy}
    M^*\equiv\underset{M}{\text{argmin}}\enskip\mathcal{M}^\mathbb{R}\left(\boldsymbol{\beta}^\mathbb{I},\{\textbf{R}_l,\boldsymbol{\mu}_l,\boldsymbol{\nu}_l\}^\mathbb{I}\right)+\mathcal{M}^\mathbb{I}\left(\boldsymbol{\beta}^\mathbb{R},\{\textbf{R}_l,\boldsymbol{\mu}_l,\boldsymbol{\nu}_l\}^\mathbb{R}\right)+\mathcal{H}^\mathbb{R}(\nu_\epsilon^\mathbb{I})+\mathcal{H}^\mathbb{I}(\nu_\epsilon^\mathbb{R})
\end{equation}
The details on the optimization algorithm used to solve Equation (\ref{eq:Real Interior Problem Effendy}) through (\ref{eq:Exterior Problem Effendy}) are discussed in the appendix.

\subsection{Construction of Credible Intervals}\label{subsec:Credible Intervals}
The details on the construction of credible intervals for $\boldsymbol{\beta}$ and $\{G_l(v)\}$ are likewise discussed in the appendix. Briefly, we use the Markov chain Monte Carlo algorithm to construct 8 independent sample chains of appropriate length. Each Monte Carlo sample corresponds to a set of parameter and distribution values drawn from the joint posterior probability density function of $\boldsymbol{\beta}$ and $\{G_l(v)\}$. We construct the 95\% credible interval by rejecting the highest and lowest 2.5\% of parameter and distribution values in the combined sample chain. 

\subsection{Comparison with Existing Algorithms}\label{subsec:Comparison with Existing Algorithms}
We compare the performance of the simplified generalized EIS inversion algorithm, hereon referred to as the gEISi algorithm, against the RR/RI algorithm, the LEVM algorithm, and the ISGP algorithm as we report them in the present work. The RR/RI algorithm is run with $M=81$. For the LEVM algorithm, the standard deviation of $\textbf{g}$ is calculated using the MATLAB command \textit{nlparci}. For the ISGP algorithm, the basis function is fixed to Gaussian, $M_H$ is set to 10, and the interior and exterior problems are solved using the non-dominated sorting genetic algorithm-II (NSGA-II) of Deb, Agrawal, Pratap, and Meyarivan \cite{Deb00}. We collect synthetic data within the angular frequency range $\omega_1={10}^{-2}$ and $\omega_J={10}^2$ with 10 data points per decade of angular frequency, with the underlying model:
\begin{equation}
    \hat{Z}(\omega)=\frac{1}{1+i\omega\exp(-2)}+\frac{1}{1+\left(i\omega\exp(2)\right)^{0.8}}
\end{equation}
The true DRT is the sum of a Dirac delta and a Cole-Cole distribution:
\begin{equation}
    G(v)=\delta(v+2)+\frac{1}{2\pi}\frac{\sin\left(0.2\pi\right)}{\cosh\left(0.8[v-2]\right)-\cos\left(0.2\pi\right)}
\end{equation}

\begin{figure*}[h!]
    \centering
    \includegraphics[scale=0.42]{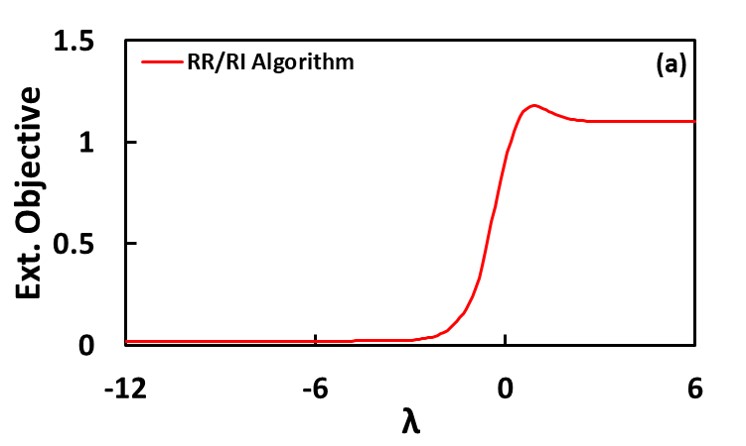}      
    \includegraphics[scale=0.42]{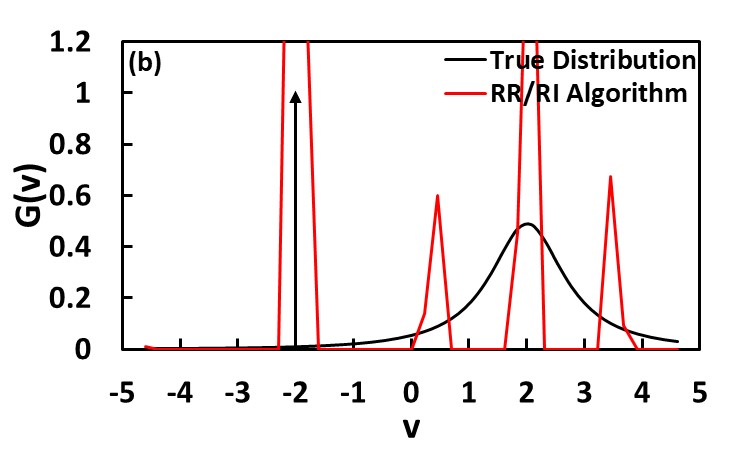}
\end{figure*}
\begin{figure*}[h!]
    \centering
    \includegraphics[scale=0.42]{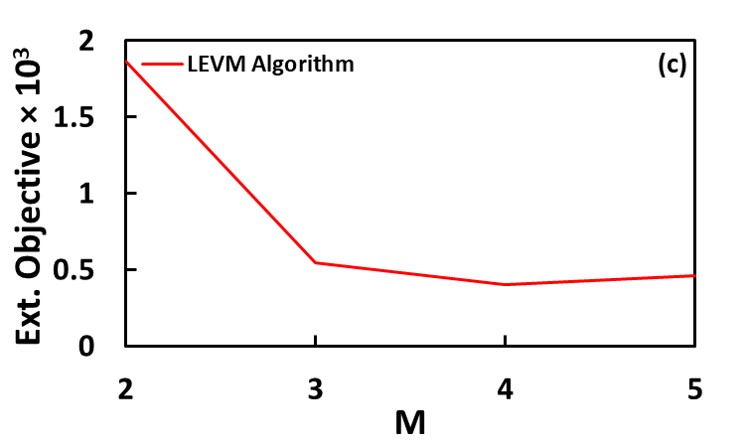}
    \includegraphics[scale=0.42]{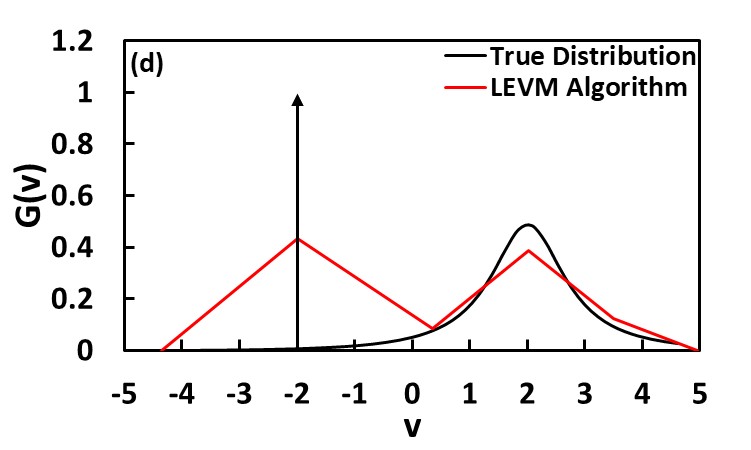}
\end{figure*}
\begin{figure*}[h!]
    \centering
    \includegraphics[scale=0.42]{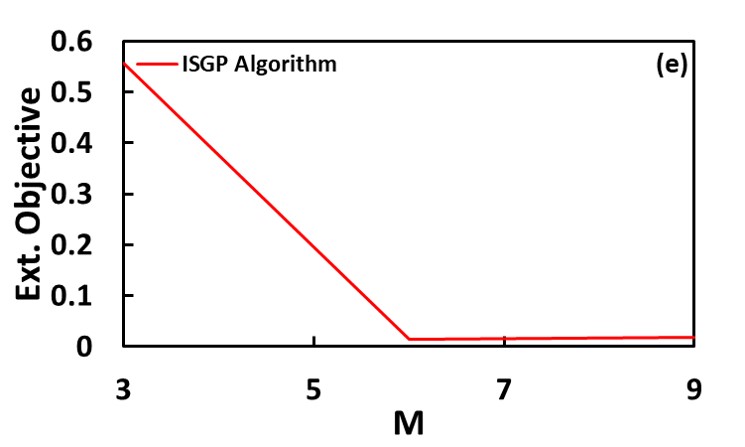}
    \includegraphics[scale=0.42]{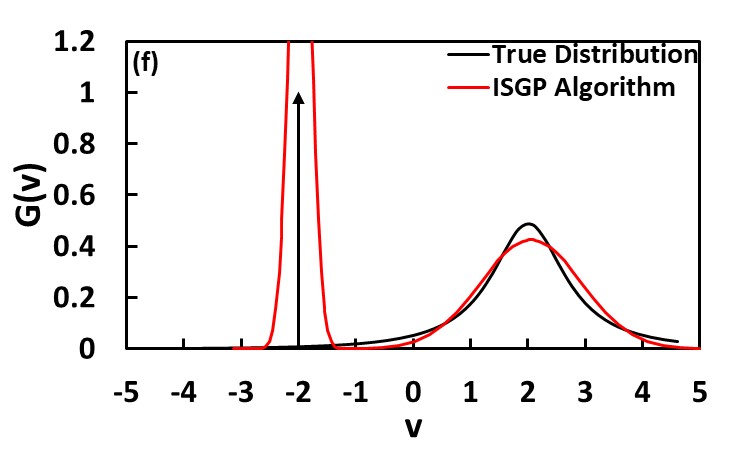}
\end{figure*}
\begin{figure*}[h!]
    \centering
    \includegraphics[scale=0.42]{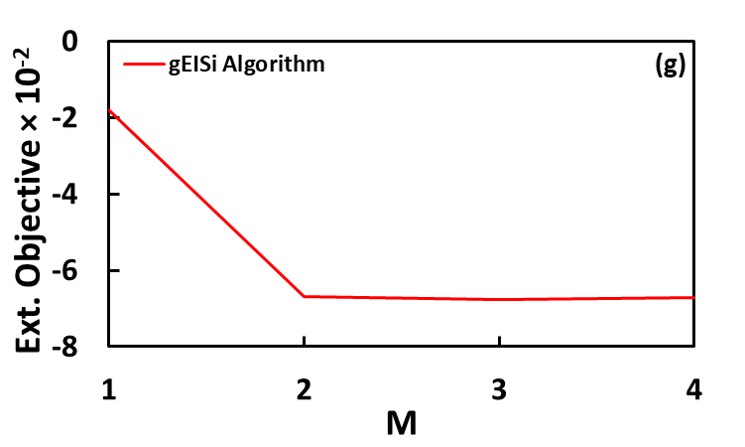}
    \includegraphics[scale=0.42]{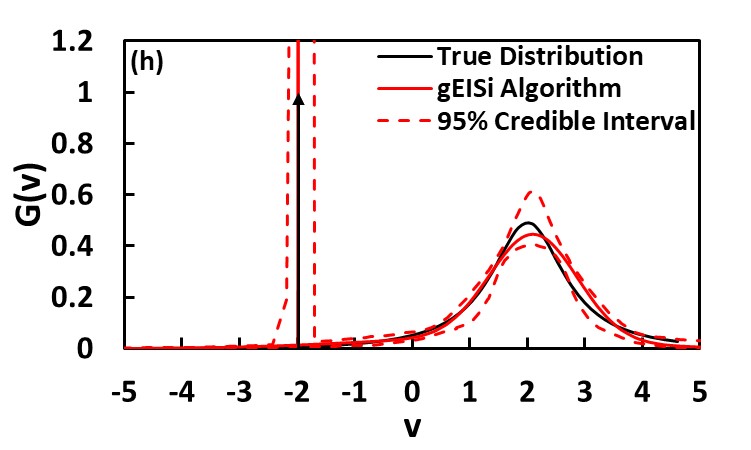}
    \caption{The objective of the exterior problem as a function of the complexity control parameter and the inversion output of (a, b) the RR/RI algorithm, (c, d) the LEVM algorithm, (e, f) the ISGP algorithm, and (g, h) the gEISi algorithm. The objective of the exterior problem and the complexity control parameter are described in (a) Section \ref{subsec:RR/RI Algorithm}, (c) Section \ref{subsec:LEVM Algorithm}, (e) Section \ref{subsec:ISGP Algorithm}, and (g) Section \ref{subsec:gEISi Algorithm}.}
    \label{fig:Final Sample Problem}
\end{figure*}

The inversion output of the four algorithms and the corresponding objective of the exterior problem as a function of the complexity control parameter are shown in Figure \ref{fig:Final Sample Problem}. The exterior problem of the RR/RI, LEVM, ISGP, and gEISi algorithm converges at $\log_{10}[\lambda^*]=-8.7$, $M^*=4$, $M^*=6$, and $M^*=3$, respectively. The RR/RI algorithm and the LEVM algorithm are unable to simultaneously reproduce the sharp and smooth features of the DRT. The failure of the RR/RI algorithm may be attributed to an inappropriate choice of complexity, while that of the LEVM algorithm arises due to the tendency for $\overline{\sigma}[g_m]$ to diverge as $g_m\xrightarrow{}{0}$, suggesting an inappropriate exterior problem. Increasing the domain of $\tau$ results in a marginal improvement in fit but does not change the qualitative aspects of the solutions. In contrast, by choosing a complexity which approximately aligns with the phenomenology described in Section \ref{sec:Design of Algorithm}, the ISGP algorithm and the gEISi algorithm are able to reproduce the Dirac delta and the Cole-Cole distribution accurately. The 95\% credible interval obtained by the gEISi algorithm likewise bounds the true distribution accurately.

\subsection{General Validation}\label{subsec:General Validation}
\begin{figure}[h!]
    \centering
    \includegraphics[scale=0.42]{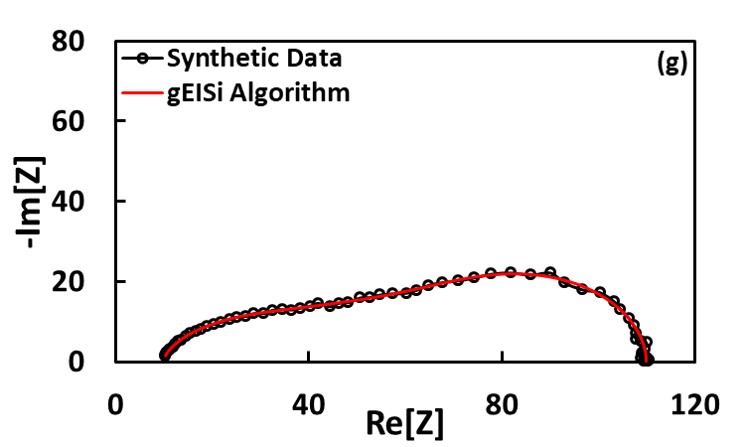}
    \includegraphics[scale=0.42]{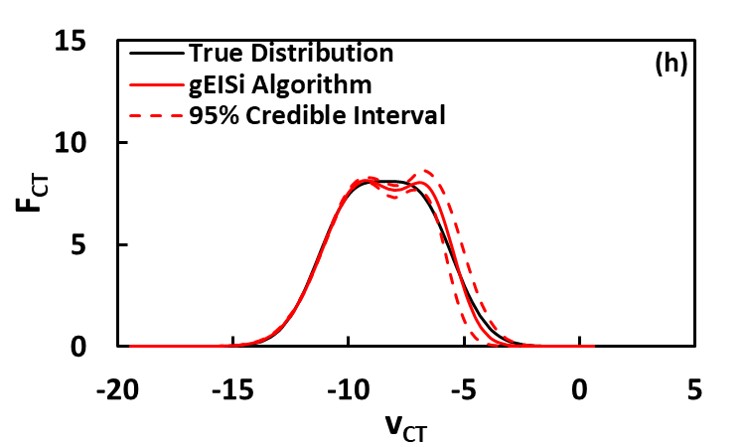}
    \includegraphics[scale=0.42]{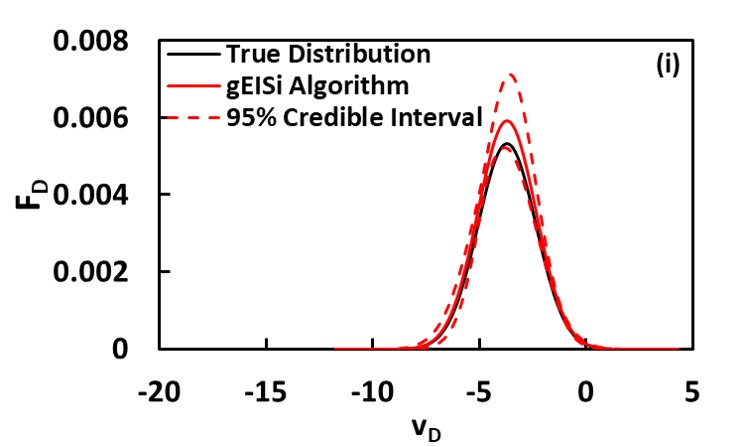}
    \caption{(a,d,g) Nyquist plot of the impedance data as well as the prediction obtained from the inversion output of the gEISi algorithm. (b,e,h) True and inverted distributions corresponding to the charge transfer process. (c,f,i) True and inverted distributions corresponding to the diffusion process. The first, second, and third row of figures correspond to the first, second, and third problems listed in Table \ref{tab:Validation Problem Parameters}, respectively.}
    \label{fig:Validation Problem}
\end{figure}

We further validate the gEISi algorithm using three generalized EIS inversion problems taking the form of Equation (\ref{eq:Distributed Diffusion}) and Equation (\ref{eq:Distributed Charge Transfer}). The distributions of each problem possess nearly overlapping peaks, which is reflected in the merging of semicircular features in Figure \ref{fig:Validation Problem} (a,d,g). Although the concept is not directly applicable here, we note that the separation between the distribution peaks is close to the resolution limit discussed by Florsch, Revil, and Camerlynck \cite{Florsch14}, which highlights the inherent difficulty of the chosen problems. The first problem illustrates an idealized situation in which the true underlying distributions are unimodal Gaussian, and are thus exactly represented by a small number of basis functions in the chosen representation. The second problem explores the effect of basis function mismatch by using unimodal Cole-Cole distributions, which cannot be exactly approximated by a finite number of Gaussian basis functions. The third problem shows the effect of increasing the modality of the true underlying distributions. 

\begin{table}[h!]
    \centering
    \begin{tabular} {|c|c|c|}
    \hline
    & $G_{CT}$ & $G_{D}$ \\
    \hline
    Problem 1 & $\frac{R_1}{\sigma_1\sqrt{2\pi}}\exp\left[-\frac{\left(v-\ln\tau_1\right)^2}{2\sigma_1^2}\right]$ & $\frac{1}{R_2\sigma_2\sqrt{2\pi}}\exp\left[-\frac{\left(v-\ln\tau_2\right)^2}{2\sigma_2^2}\right]$ \\
    Problem 2 & $\frac{R_1}{2\pi}\frac{\sin\left([1-\alpha_1]\pi\right)}{\cosh\left(\alpha_1[v-\ln\tau_1]\right)-\cos\left([1-\alpha_1]\pi\right)}$ & $\frac{1}{R_2 2\pi}\frac{\sin\left([1-\alpha_1]\pi\right)}{\cosh\left(\alpha_2[v-\ln\tau_2]\right)-\cos\left([1-\alpha_2]\pi\right)}$ \\
    Problem 3 & $\sum_{m_1=1}^2\frac{R_{m_1}}{\sigma_{m_1}\sqrt{2\pi}}\exp\left[-\frac{\left(v-\ln\tau_{m_1}\right)^2}{2\sigma_{m_1}^2}\right]$ & $\frac{1}{R_3\sigma_3\sqrt{2\pi}}\exp\left[-\frac{\left(v-\ln\tau_3\right)^2}{2\sigma_3^2}\right]$ \\
    \hline
    \end{tabular}
\end{table}

\begin{table}[h!]
    \centering
    \begin{tabular}{|c|c|c|c|c|c|}
    \hline
    & Problem 1 & & Problem 2 & & Problem 3 \\
    \hline
    $R_1$ & 50 & $R_1$ & 50 & $R_1$ & 25 \\
    $R_2$ & 50 & $R_2$ & 50 & $R_2$ & 25 \\
    $\sigma_1$ & 1.5 & $\alpha_1$ & 0.8 & $R_3$ & 50 \\
    $\sigma_2$ & 1.5 & $\alpha_2$ & 0.8 & $\sigma_1$ & 1.5 \\
    $\tau_1$ & 0.001 & $\tau_1$ & 0.001 & $\sigma_2$ & 1.5 \\
    $\tau_2$ & 0.02 & $\tau_2$ & 0.02 & $\sigma_3$ & 1.5 \\
    & & & & $\tau_1$ & 0.00005 \\
    & & & & $\tau_2$ & 0.001 \\
    & & & & $\tau_3$ & 0.02 \\
    \hline
    \end{tabular}
    \caption{True underlying distributions of the generalized EIS inversion problems under consideration and their corresponding parameter values. In all three problems, we set $R_\infty$ to 10 and the measurement error to $Z=\hat{Z}+|\hat{Z}|\times{0.005}\times(N(0,1)+iN(0,1))$.}
    \label{tab:Validation Problem Parameters}
\end{table}

We collect 10 noisy impedance measurements per decade of angular frequency between $\omega_1=10^{-2}$ and $\omega_J=10^6$, and invert the underlying distributions using the gEISi algorithm. The results are summarized in Figure \ref{fig:Validation Problem}. As shown in Figure \ref{fig:Validation Problem} (a,d,g), the impedance spectra of all three problems are reproduced satisfactorily. The true underlying distributions of the first (Figure \ref{fig:Validation Problem} (b,c)) and third (Figure \ref{fig:Validation Problem} (h,i)) problems are reproduced accurately, with the remaining difference correctly captured by the 95\% credible intervals. However, the true underlying distribution of the second problem, in particular $G_{CT}$, is not correctly bounded by the corresponding credible interval. This issue is persistent across randomly generated measurement noise profiles and stochastic optimization steps. We believe that this issue arises because the mismatch between the single Gaussian basis function used to approximate $G_D$ is fortuitously captured by the addition of a small Gaussian peak in $G_{CT}$ at $t\approx-5$. This hypothesis is supported by the large credible interval of $G_D$, which suggests that the impedance spectra effected by $G_D$ can be partially or fully accounted for by adjusting the small Gaussian peak in $G_{CT}$. The frequency with which this issue is encountered decreases rapidly with the peak-to-peak separation of $G_{CT}$ and $G_D$, and becomes nonexistent when $\log_{10}(\tau_2/\tau_1)$ exceeds 2.

The overall performance of the gEISi algorithm as applied to the distributed Randles circuit appears to be satisfactory. In all three validation problems, the numbers, sizes, and average timescales of the charge transfer and diffusion processes are correctly estimated, and the true underlying distributions are competently bounded by the 95\% credible intervals. The gEISi code, implemented in MATLAB, can be downloaded from https://github.com/suryaeff/gEISi.git.

\begin{figure}[h!]
    \centering
    \includegraphics[scale=0.42]{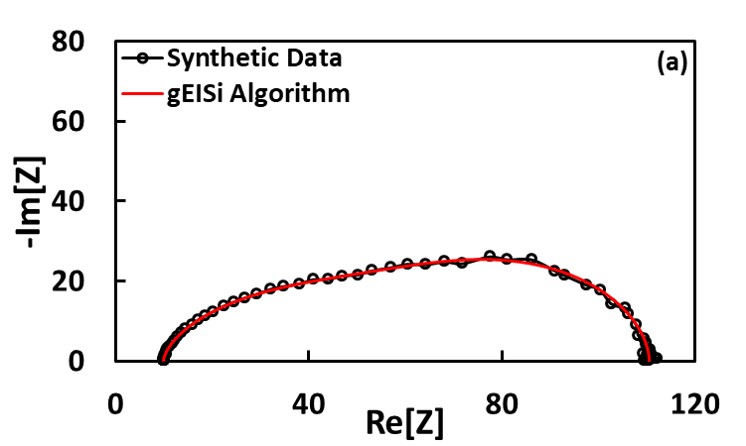}
    \includegraphics[scale=0.42]{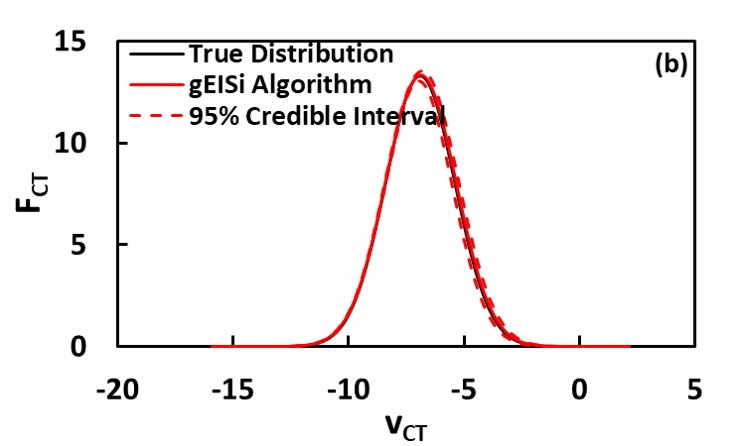}
    \includegraphics[scale=0.42]{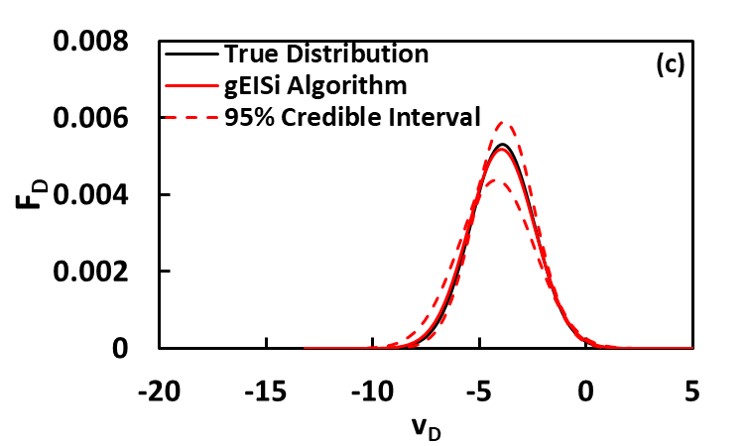}
\end{figure}
\begin{figure}[h!]
    \centering
    \includegraphics[scale=0.42]{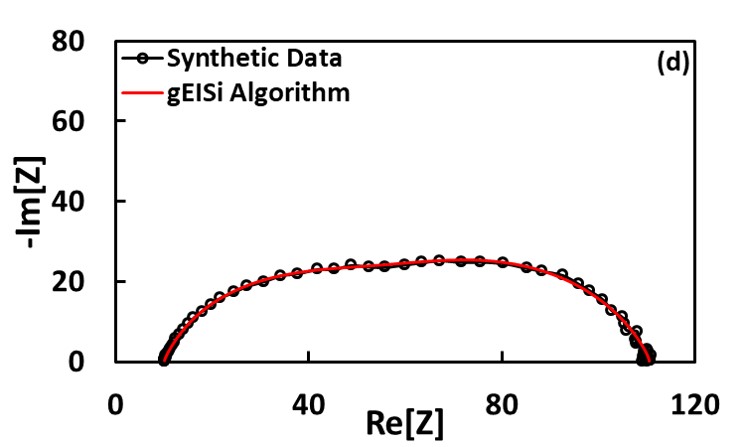}
    \includegraphics[scale=0.42]{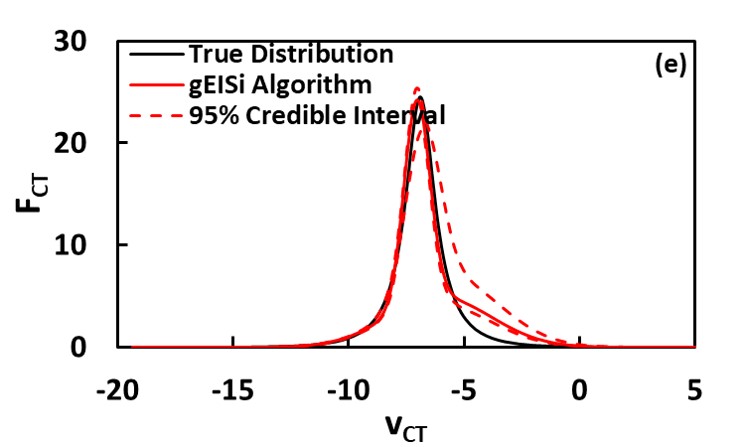}
    \includegraphics[scale=0.42]{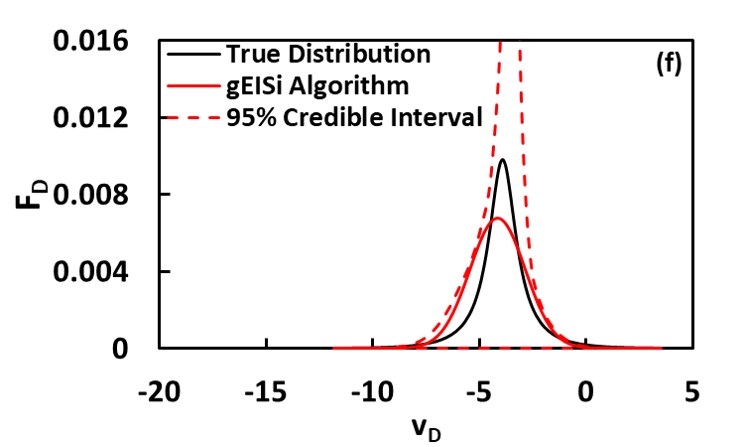}
\end{figure}

\section{Conclusion}
The present work have argued that all well-posed EIS inversion algorithms possess four common features, namely, the representation, the interior problem, the complexity control parameter, and the exterior problem. These features can be constructed and analyzed nearly independently of each other, leading to the concept of a framework for the design of EIS inversion algorithms. The ubiquity of these features has been demonstrated through a review of established DRT inversion algorithms, and are subsequently used to troubleshoot several ill-defined DRT inversion algorithms, and understand the unexpected behavior of others.

Complexity has been linked to model selection, leading to the conclusion that it should be appropriately defined in terms of the modality and the variance of the distribution. This observation, in conjunction with practical considerations on the availability of computational resource, led to the development of the gEISi algorithm, which is applicable to the generalized EIS inversion problem. The gEISi algorithm employs a floating-mesh Gaussian basis function representation, coupled to an interior problem which minimizes a misfit subject to a constraint on the control parameter, which we define as the total number of basis functions. The optimal control parameter value is determined through an exterior problem which minimizes the real-imaginary cross-validation error. The credible intervals of the parameters and distributions arising from the gEISi algorithm are estimated using Markov chain Monte Carlo algorithm. A preliminary comparison between the gEISi algorithm and several established DRT inversion algorithm suggests that it is able to overcome several limitations which has been observed in the literature. The gEISi algorithm is also validated against several generalized EIS inversion problems taking the form of distributed Randles circuit, which are impossible to analyze using existing DRT inversion algorithms. 

It should be made clear, however, that the success of the gEISi algorithm is dependent on the ability to determine the true underlying model. This allows the determination of the physicochemical parameters of the system, distributed or otherwise. In the companion paper \cite{effendy21}, we will apply the gEISi algorithm coupled to the distributed Randles circuit to analyze several experimental data sets arising from the impedance of polymer-coated steel surfaces.

\subsection*{Acknowledgement}
We gratefully acknowledge support from Dow through the University Partnership Initiative. We would like to thank Hongbo Zhao, Dimitrios Fraggedakis, and Yu Ren Zhao for their assistance in interpreting the behavior of various algorithms. We would like to thank Professor Francesco Ciucci in clarifying certain aspects of the hierarchical Bayesian algorithm. 

\section*{Appendix}
The resolution of Equation (\ref{eq:Real Interior Problem Effendy}) through (\ref{eq:Exterior Problem Effendy}) is remarkably difficult, because the model is input by the end-user, and therefore not known a priori. In such cases, it is prudent to use a combination of optimization algorithms; in the present work, three optimization algorithms are used, namely, simulated annealing, constrained trust region, and constrained pattern search. These will be discussed individually in the subsequent paragraphs.

\begin{figure*}[h!]
    \centering
    \includegraphics[scale=0.42]{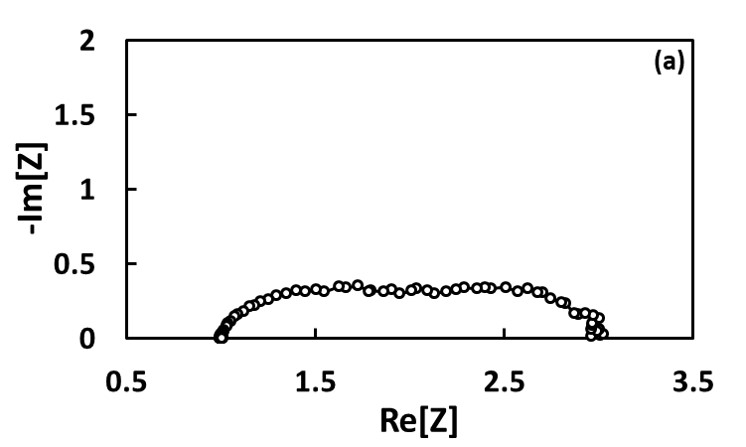}
    \includegraphics[scale=0.42]{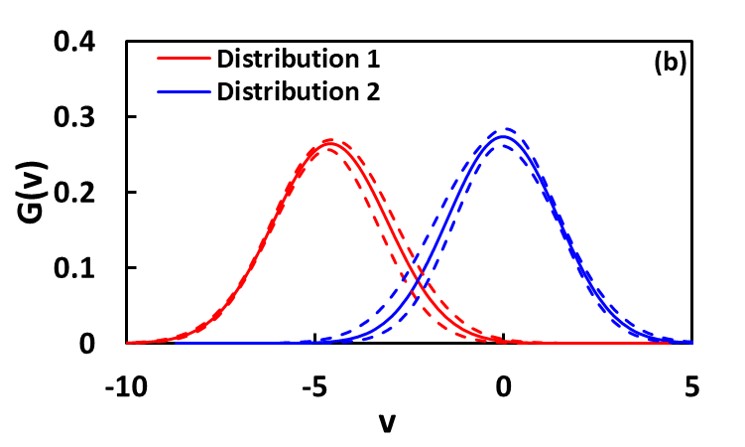}
    \includegraphics[scale=0.42]{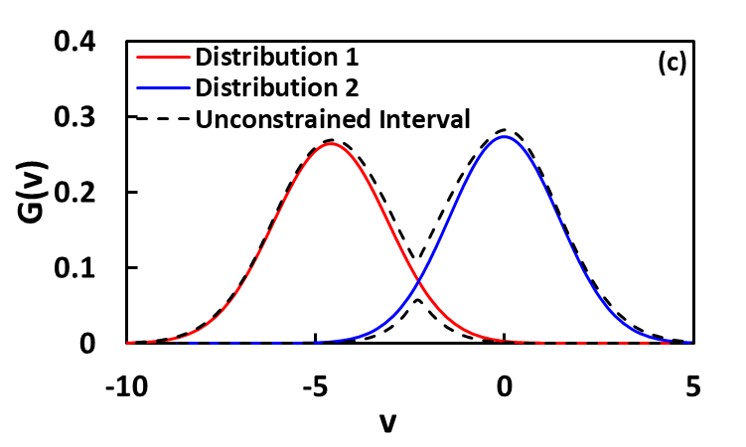}
    \caption{The impedance data in (a) is inverted using a model consisting of two distributed relaxation processes, yielding the two distributions shown in (b). The dashed lines of corresponding color is the 95\% credible interval of the distributions. (c) In the absence of Equation (\ref{eq:Interchangeability Constraint 2}), the 95\% credible interval of the distributions are indistinguishable and overestimated.}
    \label{fig:Interchangeability}
\end{figure*}

The construction of credible intervals for $\boldsymbol{\beta}$ and $\{G_l(v)$ necessitated the addition of constraints on the interchangeability of basis functions. These constraints take the form:
\begin{align}\label{eq:Interchangeability Constraint 1}
    \mu_{m_l}&\leq\mu_{m_l+1}\\\label{eq:Interchangeability Constraint 2}
    \mu_{m_l=M_l}&\leq\mu_{m_{l+1}=1}
\end{align}
The first constraint states that the means of the basis functions in the $l^\text{th}$ distribution must be increasing. In the absence of this constraint, basis functions can interchange, leading to an inflated estimate of the credible intervals for $\{\textbf{R}_l,\boldsymbol{\mu}_l,\boldsymbol{\nu}_l\}$. The second constraint states that the highest mean of the $l^\text{th}$ distribution must be lower than the lowest mean of the $(l+1)^\text{th}$ distribution. In the absence of this constraint, basis functions can interchange among different distributions, leading to an inflated estimate of the credible intervals for $\{\textbf{R}_l,\boldsymbol{\mu}_l,\boldsymbol{\nu}_l\}$ and $\{G_l(v)\}$. The concept of interchangeability among different basis function is illustrated in Figure \ref{fig:Interchangeability}. It is also convenient to set a convention whereby $\{G_l(v)\}$ is nonnegative:
\begin{equation}\label{eq:Mass Constraint}
    R_{m_l}\geq{0}
\end{equation}
This arises from the assumption that the end-user knows with certainty that the form of the model is correct up to a sign. Trial-and-error suggests that this constraint is helpful in reducing the computational time needed to solve the exterior problem. 

The initial guess for the interior problem at each value of $M$ is generated from the previous interior problem by decomposing the previous optimal fit using a moment-matching equation. The exterior problem, i.e., Equation (\ref{eq:Exterior Problem Effendy}) is solved by sequential enumeration. The gEISi code, implemented in MATLAB, can be downloaded from the supporting information. 

\subsubsection*{Simulated Annealing}
For brevity, let $\boldsymbol{\gamma}$ be a vector of all parameter values arising from the chosen Gaussian representation, with $\gamma_g$ as the $g^\text{th}$ element of $\boldsymbol{\gamma}$. The size of $\boldsymbol{\gamma}$ is $K+3M+1$, since the point parameters $\boldsymbol{\beta}$, distributed parameters $\{G_l(v)\}$, and the log-variance of the measurement error $\nu_\epsilon$ give rise to K, 3M, and 1 parameter each. Let $\boldsymbol{\gamma}^*$ be the maximum likelihood estimate of $\boldsymbol{\gamma}$, $\boldsymbol{\gamma}^{(p)}$ be the proposed step of the simulated annealing, and let $\boldsymbol{\gamma}^{(n,g)}$ be the ${(n,g)}^\text{th}$ step of the simulated annealing. Finally, let $N_\text{s}$ be the effective number of successful iterations of simulated annealing.  

The simulated annealing code requires as inputs the perturbation size $\Delta\boldsymbol{\gamma}$ and the decorrelation length $N_\text{d}$. The derivation of $\Delta\boldsymbol{\gamma}$ and $N_\text{d}$ will be discussed in a subsequent section. The pseudocode of the simulated annealing takes the form:
\begin{align*}
    &\text{Set }\boldsymbol{\gamma}^*\text{ to }\boldsymbol{\gamma}^{(n,g)}\\
    &\text{Set }N_\text{s}\text{ to }1\\
    &\text{While }N_\text{s}\text{ is less than }N_\text{d},\\
    &\qquad\text{Set }\boldsymbol{\gamma}^{(n+1,g)}\text{ to }\boldsymbol{\gamma}^{(n,g)}\\
    &\qquad\text{For each element in }\boldsymbol{\gamma}\text{, perform the update:}\\
    &\qquad\qquad\gamma^{(p)}_{g'}=\begin{cases}
    \gamma^{(n,g)}_{g'}+N(0,1)\Delta\gamma_g\text{ for the element under consideration, and}\\
    \gamma^{(n,g)}_{g'}\text{ otherwise}
    \end{cases}\\
    &\qquad\qquad p_\mu=\begin{cases}
    1\text{ if Equation (\ref{eq:Interchangeability Constraint 1}) and Equation (\ref{eq:Interchangeability Constraint 2}) are satisfied}\\
    0\text{ otherwise}
    \end{cases}\\
    &\qquad\qquad p_R=\begin{cases}
    1\text{ if Equation (\ref{eq:Mass Constraint}) is satisfied}\\
    0\text{ otherwise}
    \end{cases}\\
    &\qquad\qquad p=\min\left[1,p_\mu p_R \exp\left(-\frac{N_\text{s}}{N_\text{d}}\frac{\mathcal{M}(\boldsymbol{\gamma}^{(p)})+\mathcal{H}(\boldsymbol{\gamma}^{(p)})-\mathcal{M}(\boldsymbol{\gamma}^{(n,g)})-\mathcal{H}(\boldsymbol{\gamma}^{(n,g)})}{2}\right)\right]\\
    &\qquad\qquad\boldsymbol{\gamma}^{(n+1,g)}=\begin{cases}
    \boldsymbol{\gamma}^{(p)}\text{ if }U[0,1]<p\\
    \boldsymbol{\gamma}^{(n,g)}\text{ otherwise}
    \end{cases}\\
    &\qquad\qquad\text{If }\mathcal{M}(\boldsymbol{\gamma}^{(n+1,g)})+\mathcal{H}(\boldsymbol{\gamma}^{(n+1,g)})<\mathcal{M}(\boldsymbol{\gamma}^{(n,g)})+\mathcal{H}(\boldsymbol{\gamma}^{(n,g)})\text{,}\\
    &\qquad\qquad\qquad\boldsymbol{\gamma}^*=\boldsymbol{\gamma}^{(n+1,g)}\\
    &\qquad\qquad\qquad N_\text{s}=N_\text{s}+\frac{1}{K+3M+1}\\
    &\qquad\qquad\qquad\text{Update }\Delta\boldsymbol{\gamma}\\
    &\qquad\qquad\text{End}\\
    &\qquad\text{End}\\
    &\text{End}
\end{align*}
Here $N(0,1)$ and $U[0,1]$ are random numbers generated from the standard normal and standard uniform distribution, respectively, and $p$ is the acceptance probability of the proposed step. In effect, this pseudocode performs Markov chain Monte Carlo under an increasingly relaxed condition as $N_\text{s}$ approaches 0. The Monte Carlo iterations are done one parameter at a time, with each accepted step accounting for a fraction of an effective number of successful iterations. 

\subsubsection*{Perturbation Size and Decorrelation Length}
For the simulated annealing to succeed, the parameter space explored by the Monte Carlo steps should encompass the maximum likelihood estimate. In this section, we derive approximate expressions for the decorrelation length and the perturbation size, defined as the number and size of successful Monte Carlo steps needed to explore a parameter space encompassing the maximum likelihood estimate with confidence $1-\alpha$, respectively. 

Let $\chi^2(\boldsymbol{\gamma})\equiv\mathcal{M}(\boldsymbol{\gamma})+\mathcal{H}(\boldsymbol{\gamma})$. In a regular Markov chain Monte Carlo algorithm, we have:
\begin{equation}
    p=\min\left[1,\exp\left(-\frac{\chi^2(\boldsymbol{\gamma}^{(p)})-\chi^2(\boldsymbol{\gamma}^{(n,g)})}{2}\right)\right]
\end{equation}
We want $p$ to be of reasonable size; excessively small $p$ leads to excessively high computational time, while excessively large $p$ leads to small, ineffective steps. Setting $p=\frac{1}{e}$, and evaluating $\boldsymbol{\gamma}^{(n,g)}$ at the maximum likelihood estimate,
\begin{equation}\label{eq:Allowed Difference}
    \chi^2(\boldsymbol{\gamma}^{(p)})-\chi^2(\boldsymbol{\gamma}^*)=2
\end{equation}
Let $\textbf{H}$ be the Hessian of $\chi^2$ evaluated at the maximum likelihood estimate. Then around the maximum likelihood estimate, we can approximate $\chi^2$ as:
\begin{equation}\label{eq:Hessian Approximation}
    \chi^2(\boldsymbol{\gamma}^{(p)})\approx\chi^2(\boldsymbol{\gamma}^*)+\frac{1}{2}(\boldsymbol{\gamma}^{(p)}-\boldsymbol{\gamma}^*)^\text{T}\textbf{H}(\boldsymbol{\gamma}^{(p)}-\boldsymbol{\gamma}^*)
\end{equation}
Since the Monte Carlo simulation is done one parameter at a time, only the diagonal elements of the Hessian matters. We assign the allowed difference in Equation (\ref{eq:Allowed Difference}) to the diagonal elements:
\begin{equation}
    \frac{1}{2}\Delta\gamma_g^2H_{g,g}\approx{2}
\end{equation}
By evaluating the diagonal elements of the Hessian at the current maximum likelihood estimate, provided that the maximum likelihood estimate does not fluctuate excessively throughout the simulated annealing process, a consistent and reasonable perturbation size can be obtained. 

Next, we consider the decorrelation length. Our strategy is to obtain the size of the parameter space encompassing $1-\alpha$ of the probability mass relative to the perturbation size, and calculate the decorrelation length by approximating successful Monte Carlo steps as random walk. 

Let $\varepsilon^\mathbb{R}_j\equiv\frac{\text{re}[Z_j-\hat{Z}_j])}{\text{abs}(Z_j)}$ and $\varepsilon^\mathbb{I}_j\equiv\frac{\text{im}[Z_j-\hat{Z}_j])}{\text{abs}(Z_j)}$. We want to obtain some limiting error $\varepsilon'$ which encompasses $1-\alpha$ of the probability mass:
\begin{equation}
    \Pi_j\int_{-\varepsilon'}^{\varepsilon'}{P(\varepsilon^\mathbb{R}_j)d\varepsilon^\mathbb{R}_j}\int_{-\varepsilon'}^{\varepsilon'}{P(\varepsilon^\mathbb{I}_j)d\varepsilon^\mathbb{I}_j}=1-\alpha
\end{equation}
For a Gaussian independently and identically distributed error, we have, as a conservative estimate,
\begin{equation}
    \int_{-\varepsilon'}^{\varepsilon'}{\frac{1}{\sqrt{2\pi}\exp(\nu_\epsilon/2)}\exp\left[-\frac{\varepsilon^2_j}{2\exp(\nu_\epsilon)}\right]d\varepsilon_j}=(1-\alpha)^\frac{1}{2J}
\end{equation}
Equivalently,
\begin{equation}\label{eq:Solution to Error Size}
    \varepsilon'=\exp(\nu_\epsilon/2)\text{norminv}\left[\frac{1}{2}\left(1+[1-\alpha]^\frac{1}{2J}\right)\right]
\end{equation}
Here $\text{norminv}[\cdot]$ is the standard normal inverse operator. Substituting the limiting error into Equation (\ref{eq:Hessian Approximation}), and neglecting the difference in the hyperprior terms,
\begin{equation}
    \frac{1}{2}(\Delta\boldsymbol{\gamma})_\text{max}^\text{T}\textbf{H}(\Delta\boldsymbol{\gamma})_\text{max}\approx{2\sum_j\left[\frac{\varepsilon_j'}{\exp(\nu_\epsilon/2)}\right]^2-\mathcal{M}(\boldsymbol{\gamma}^*)}
\end{equation}
Here $(\Delta\boldsymbol{\gamma})_\text{max}$ is the size of the parameter space corresponding to the limiting error. Substituting the expression for limiting error,
\begin{equation}
    \frac{1}{2}(\Delta\boldsymbol{\gamma})_\text{max}^\text{T}\textbf{H}(\Delta\boldsymbol{\gamma})_\text{max}\approx{2J\text{norminv}\left[\frac{1}{2}\left(1+[1-\alpha]^\frac{1}{2J}\right)\right]^2-\mathcal{M}(\boldsymbol{\gamma}^*)}
\end{equation}
We neglect the off-diagonal elements of the Hessian, and split the right-hand side of the equation among the diagonal elements:
\begin{equation}
    \frac{1}{2}(\Delta\gamma_g)_\text{max}^2H_{g,g}\approx\frac{2J\text{norminv}\left[\frac{1}{2}\left(1+[1-\alpha]^\frac{1}{2J}\right)\right]^2-\mathcal{M}(\boldsymbol{\gamma}^*)}{K+3M+1}
\end{equation}

It remains to compare the perturbation size to the size of the parameter space corresponding to the limiting error. Taking each successful Monte Carlo step as a random walk, and approximating the misfit $\mathcal{M}(\boldsymbol{\gamma}^*)$ with the idealized value of $2J$,
\begin{equation}\label{eq:Decorrelation Size}
    N_\text{d}=\left[\frac{2(\Delta\gamma_g)_\text{max}}{\Delta\gamma_g}\right]^2\approx\frac{2(2J)\text{norminv}\left[\frac{1}{2}\left(1+[1-\alpha]^\frac{1}{2J}\right)\right]^2-2(2J)}{K+3M+1}
\end{equation}
In practice, the Monte Carlo steps do not fully resemble a random walk, since they are directed towards $\boldsymbol{\gamma}^*$. To account for this drift, we scale $N_\text{d}$ according to the dimensionality of the parameter space:
\begin{equation}
    N_\text{d}\approx\frac{2(2J)\text{norminv}\left[\frac{1}{2}\left(1+[1-\alpha]^\frac{1}{2J}\right)\right]^2-2(2J)}{\sqrt{K+3M+1}}
\end{equation}
We also drop the term $-2(2J)$ to allow for a conservative estimate. Equation (\ref{eq:Decorrelation Size}) has been derived assuming that the full data set has been used, corresponding to $2J$ degrees of freedom. For Equation (\ref{eq:Real Interior Problem Effendy}) and Equation (\ref{eq:Imaginary Interior Problem Effendy}), every instance of $2J$ should be replaced with $J+1$. 

\subsubsection*{Constrained Trust Region}
MATLAB's default constrained trust region algorithm \textit{fmincon} is used to solve Equation (\ref{eq:Real Interior Problem Effendy}) and Equation (\ref{eq:Imaginary Interior Problem Effendy}), setting the maximum number of function evaluations and the maximum number of iterations to $100(K+3M+1)^2$. The usual constraints on interchangeability and mass of basis functions apply. Constrained trust region is used to refine the solution obtained using simulated annealing. 

\subsubsection*{Constrained Pattern Search}
The execution of constrained pattern search is identical to the constrained trust region, except done one parameter at a time, and using the default settings of \textit{fmincon}. The constrained pattern search is performed whenever $\Delta{\gamma}$ is updated in the simulated annealing pseudocode. Trial-and-error suggests that the constrained pattern search is a quick and effective method of accelerating the convergence of the simulated annealing, without sacrificing the stochasticity of the simulated annealing. 

\subsection*{Interior Problem}
The simulated annealing, constrained trust region, and constrained pattern search are combined to solve the interior problem. Assume that a rough initial guess of the interior problem $\boldsymbol{\gamma}^{(0)}$ exists, and refine this initial guess using the optimization problem:
\begin{align}\label{eq:Initial Interior Guess}
    \boldsymbol{\beta}^\mathbb{O},\{\textbf{R}_l,\boldsymbol{\mu}_l,\boldsymbol{\nu}_l\}^\mathbb{O},\nu_\epsilon^\mathbb{O}&\equiv\underset{\boldsymbol{\beta},\{\textbf{R}_l,\boldsymbol{\mu}_l,\boldsymbol{\nu}_l\},\nu_\epsilon}{\text{argmin}}\enskip\mathcal{M}^\mathbb{O}+\mathcal{H}^\mathbb{O}\\
    \mathcal{M}^\mathbb{O}&\equiv\frac{\left\lVert{(\textbf{Z}-\hat{\textbf{Z}})\,\oslash\,\text{abs}[\textbf{Z}]}\right\rVert_2^2}{\exp(\nu_\epsilon)}\\
    \mathcal{H}^\mathbb{O}&\equiv\frac{(\mu_\epsilon-\nu_\epsilon)^2}{\sigma_\epsilon^2}+\nu_\epsilon(2J)
\end{align}
The superscript $\mathbb{O}$ denotes that the parameters are obtained using both the imaginary and the real parts of the data set. The interior problem code also requires as input the number of basis functions used to represent each distribution $\{M_l\}$, also called the modality or the case under consideration. The derivation of $\boldsymbol{\gamma}^{(0)}$ and $\{M_l\}$ will be discussed in a subsequent section. Let $X_\text{val}$ be the cross-validation error. 

The pseudocode of the interior problem takes the form:
\begin{align*}
    &\text{Set }X_\text{val}\text{ to }0\\
    &\text{Solve Equation (\ref{eq:Initial Interior Guess}) for the case under consideration, with sub-steps:}\\
    &\qquad\text{Perform simulated annealing with initial guess }\boldsymbol{\gamma}^{(0)}\text{, and set }\boldsymbol{\gamma}^\mathbb{O}\text{ to the output}\\
    &\qquad\text{Perform constrained trust region with initial guess }\boldsymbol{\gamma}^\mathbb{O}\text{, and set }\boldsymbol{\gamma}^\mathbb{O}\text{ to the output}\\
    &\text{Solve Equation (\ref{eq:Real Interior Problem Effendy}) for the case under consideration, with sub-steps:}\\
    &\qquad\text{Perform simulated annealing with initial guess }\boldsymbol{\gamma}^\mathbb{O}\text{, and set }\boldsymbol{\gamma}^\mathbb{R}\text{ to the output}\\
    &\qquad\text{Perform constrained trust region with initial guess }\boldsymbol{\gamma}^\mathbb{R}\text{, and set }\boldsymbol{\gamma}^\mathbb{R}\text{ to the output}\\
    &X_\text{val}=\mathcal{M}^\mathbb{I}(\boldsymbol{\gamma}^\mathbb{R})+\mathcal{H}^\mathbb{I}(\boldsymbol{\gamma}^\mathbb{R})\\
    &\text{Solve Equation (\ref{eq:Imaginary Interior Problem Effendy}) for the case under consideration, with sub-steps:}\\
    &\qquad\text{Perform simulated annealing with initial guess }\boldsymbol{\gamma}^\mathbb{O}\text{, and set }\boldsymbol{\gamma}^\mathbb{I}\text{ to the output}\\
    &\qquad\text{Perform constrained trust region with initial guess }\boldsymbol{\gamma}^\mathbb{I}\text{, and set }\boldsymbol{\gamma}^\mathbb{I}\text{ to the output}\\
    &X_\text{val}=X_\text{val}+\mathcal{M}^\mathbb{R}(\boldsymbol{\gamma}^\mathbb{I})+\mathcal{H}^\mathbb{R}(\boldsymbol{\gamma}^\mathbb{I})
\end{align*}

\subsection*{Sequential Enumeration}
The argument of the exterior problem ($M$) is an integer. We thus solve the exterior problem by sequentially increasing $M$, and then generating new cases and rough initial guesses for the resulting interior problem. The generation of new cases and rough initial guesses will be discussed in subsequent sections. Let $X_\text{val}^*$ be the current optimal cross-validation error, $X_\text{prev}^*$ be the previous optimal cross-validation error, $\boldsymbol{\gamma}^*$ be the current optimal maximum likelihood estimate, $\boldsymbol{\gamma}^*_\text{prev}$ be the previous optimal maximum likelihood estimate, $\{M_l\}^*$ be the current optimal modality, $\{M_l\}^*_\text{prev}$ be the previous optimal modality, and $\{\{M_l\}\}$ be the set of cases under consideration.

The pseudocode has two parts, corresponding to $M=L$, i.e., the case where each distribution is approximated by 1 basis function, and $M>L$. The pseudocode for $M=L$ takes the form:
\begin{align*}
    &\text{Solve the interior problem for }\{M_l\}=\{1,...,1\}\\
    &\text{Set }X_\text{val}^*\text{ to }X_\text{val}\\
    &\text{Set }\boldsymbol{\gamma}^*\text{ to }\boldsymbol{\gamma}^\mathbb{O}\\
    &\text{Set }\{M_l\}^*\text{ to }\{1,...,1\}
\end{align*}
This pseudocode requires a true initial guess $\boldsymbol{\gamma}^{(0)}$ for the interior problem, which is provided by the end-user.

The pseudocode for $M>L$ takes the form:
\begin{align*}
    &\text{While }X^*_\text{prev}-X^*_\text{val}>2\alpha(2J),\\
    &\qquad\text{Set }X_\text{prev}^*\text{ to }X_\text{val}^*\\
    &\qquad\text{Set }\boldsymbol{\gamma}^*_\text{prev}\text{ to }\boldsymbol{\gamma}^*\\
    &\qquad\text{Set }\{M_l\}^*_\text{prev}\text{ to }\{M_l\}^*\\
    &\qquad\text{Generate }\{\{M_l\}\}\text{ from }\{M_l\}^*_\text{prev}\\
    &\qquad\text{For each element in }\{\{M_l\}\}\text{, perform:}\\
    &\qquad\qquad\text{Generate }\boldsymbol{\gamma}^{(0)}\text{ from }\boldsymbol{\gamma}^*_\text{prev}\text{ consistent with the case under consideration}\\
    &\qquad\qquad\text{Solve the interior problem for the case under consideration}\\
    &\qquad\qquad\text{If }X_\text{val}<X_\text{val}^*,\\
    &\qquad\qquad\qquad X_\text{val}^*=X_\text{val}\\
    &\qquad\qquad\qquad\boldsymbol{\gamma}^*=\boldsymbol{\gamma}^\mathbb{O}\\
    &\qquad\qquad\qquad\text{Set }\{M_l\}^*\text{ to the case under consideration}\\
    &\qquad\qquad\text{End}\\
    &\qquad\text{End}\\
    &\text{End}
\end{align*}
At every iteration of the while loop, the control parameter $M=\sum_l{M_l}$ increases by 1. The code compares the current optimal cross-validation error against the previous optimal cross-validation error, and allows the algorithm to proceed only if the reduction in cross-validation error exceeds $2\alpha(2J)$, where $\alpha$ is the approximate probability mass left unexplored by the simulated annealing. Note that $2\alpha$ arises from a conservative estimate of the confidence of the optimization problem, while $2J$ is the idealized value of the misfit. This threshold accounts for incomplete optimization and model mismatch; since the present work uses only synthetic data, which does not experience model mismatch, $\alpha$ can be set to an arbitrarily low value. We have opted to set $\alpha=0.01$ in Section \ref{subsec:Comparison with Existing Algorithms}, which is effectively vanishing. To provide a sense of scale, with regards to Figure \ref{fig:Final Sample Problem} (d), $X^*_\text{prev}-X^*_\text{val}$ is of the order of 100, while $2\alpha(2J)$ with $\alpha=0.01$ is in the order of 1.

More generally, by calibration against the experimental data used in the companion paper, as well as other experimental data generated in our laboratory, $\alpha$ is set to a more reasonable value of 0.1. For direct comparability with the analysis published in our companion paper \cite{effendy21}, in Section \ref{subsec:General Validation}, we set $\alpha$ to 0.1.

\subsection*{Moment Matching}
The pseudocode for the interior problem requires the generation of a rough initial guess $\boldsymbol{\gamma}^{(0)}$ from the previous optimal maximum likelihood estimate $\boldsymbol{\gamma}^*_\text{prev}$. This is done by increasing the modality to match the case under consideration, and then matching the first two moments of $\boldsymbol{\gamma}^{(0)}$ and $\boldsymbol{\gamma}^*_\text{prev}$. For clarity, recall that $\{M_l\}$ is the case under consideration, $\{M_l\}^*_\text{prev}$ is the previous optimal modality, and $\boldsymbol{\gamma}$ is composed of the point parameters $\boldsymbol{\beta}$, the distribution masses $\{\textbf{R}_l\}$, the distribution means $\{\boldsymbol{\mu}_l\}$, and the distribution log-variances $\{\boldsymbol{\nu}\}_l$. Let $M^*_{l,\text{prev}}$ be the $l^\text{th}$ element of $\{M_l\}^*_\text{prev}$, and let $R_{m_l,\text{prev}}^*$, $\mu_{m_l,\text{prev}}^*$, and $\nu_{m_l,\text{prev}}^*$ be the $m_l^\text{th}$ distribution mass, mean, and log-variance of the $l^\text{th}$ distribution, respectively.

The pseudocode of the moment-matching algorithm takes the form:
\begin{align*}
    &\text{Set }\boldsymbol{\beta}^{(0)}\text{ to }\boldsymbol{\beta}^*\\
    &\text{For each element in }\{\textbf{R}_l\}^{(0)},\\
    &\qquad R_{m_l}=\begin{cases}
    \frac{R_{m_l,\text{prev}}^*}{2}\text{ if }m_l=1\\
    \frac{R_{m_l-1,\text{prev}}^*+R_{m_l,\text{prev}}^*}{2}\text{ if }2\leq{m_l}\leq{M_l}\\
    \frac{R_{m_l-1,\text{prev}}^*}{2}\text{ if }m_l=M_l
    \end{cases}\\
    &\text{For each element in }\{\boldsymbol{\mu}_l\}^{(0)},\\
    &\qquad\mu_{m_l}=\begin{cases}
    \mu_{m_l,\text{prev}}^*-\exp\left[\nu_{m_l,\text{prev}}^*/2\right]\text{ if }m_l=1\\
    \frac{R_{m_l-1,\text{prev}}^*\left(\mu_{m_l-1,\text{prev}}^*+\exp\left[\nu_{m_l-1,\text{prev}}^*/2\right]\right)+R_{m_l,\text{prev}}^*\left(\mu_{m_l,\text{prev}}^*-\exp\left[\nu_{m_l,\text{prev}}^*/2\right]\right)}{R_{m_l-1,\text{prev}}^*+R_{m_l,\text{prev}}^*}\text{ if }2\leq{m_l}\leq{M_l}\\
    \mu_{m_l-1,\text{prev}}^*+\exp\left[\nu_{m_l-1,\text{prev}}^*/2\right]\text{ if }m_l=M_l
    \end{cases}\\
    &\text{Set all elements in }\{\boldsymbol{\nu}_l\}^{(0)}\text{ to }\frac{\text{abs}[\ln(\omega_J/\omega_1)]}{J-1}
\end{align*}

\subsection*{Cases Under Consideration}
The cases under consideration $\{\{M_l\}\}$ is generated by increasing each element of $\{M_l\}^*_\text{prev}$ independently. For example, if $\{M_l\}^*_\text{prev}=\{1,2,2\}$, then $\{\{M_l\}\}=\{\{2,2,2\},\{1,3,2\},\{1,2,3\}\}$; we will then proceed to consider these three cases in the pseudocode of the exterior problem. Trial-and-error suggests this heuristic works well over a wide variety of EIS inversion problems.

\subsection*{Construction of Credible Intervals}
The construction of credible intervals for $\boldsymbol{\beta}$ and $\{G_l(v)\}$ is done using Markov chain Monte Carlo performed one parameter at a time. The execution of the algorithm is identical to simulated annealing, with the relaxation term removed:
\begin{align*}
    \exp\left(-\frac{N_\text{s}}{N_\text{d}}\frac{\mathcal{M}(\boldsymbol{\gamma}^{(p)})+\mathcal{H}(\boldsymbol{\gamma}^{(p)})-\mathcal{M}(\boldsymbol{\gamma}^{(n,g)})-\mathcal{H}(\boldsymbol{\gamma}^{(n,g)})}{2}\right)\xrightarrow{\text{replace}}\\
    \qquad\exp\left(-\frac{\mathcal{M}(\boldsymbol{\gamma}^{(p)})+\mathcal{H}(\boldsymbol{\gamma}^{(p)})-\mathcal{M}(\boldsymbol{\gamma}^{(n,g)})-\mathcal{H}(\boldsymbol{\gamma}^{(n,g)})}{2}\right)
\end{align*}
In principle, the most efficient implementation of the Monte Carlo algorithm consists of a single chain of length $\gg{N_\text{d}}$. However, due to the various approximations taken in the derivation of $N_\text{d}$, we cannot interpret it as more than a competent approximation for the decorrelation length. Trial-and-error suggests that better credible intervals are obtained by obtaining multiple independent Monte Carlo chains of length $N_\text{d}$. Simulated annealing is used to decorrelate successive runs of Markov chain Monte Carlo algorithm. 

It is useful to reiterate some of the definitions used in simulated annealing. Let $\boldsymbol{\gamma}^{(n,g)}$ be the $(n,g)^\text{th}$ step of the algorithm, with $n$ as the index of Monte Carlo sample and $g$ as the index of parameter under consideration. After every parameter has been considered once, the index $n$ increases, and a single Monte Carlo sample is obtained. We define $\boldsymbol{\gamma}^{(n)}\equiv\boldsymbol{\gamma}^{(n,1)}$ to be the $n^\text{th}$ Monte Carlo sample of the algorithm. Let $\{\boldsymbol{\gamma}^{(n)}\}$ be the collection of samples arising from a single run of the Monte Carlo algorithm, and let $\{\boldsymbol{\gamma}\}$ be the collection of samples arising from all runs of the Monte Carlo algorithm. The pseudocode for the construction of credible intervals takes the form:
\begin{align*}
    &\text{Set }N_\text{MS}\text{ to }1\\
    &\text{Set }\boldsymbol{\gamma}^{(n,g)}\text{ to }\gamma^*\\
    &\text{While }N_\text{MS}\leq{8},\\
    &\qquad\text{Perform Markov chain Monte Carlo with initial guess }\boldsymbol{\gamma}^{(n,g)}\text{, and collect }\{\boldsymbol{\gamma}^{(n)}\}\\
    &\qquad\text{Append }\{\boldsymbol{\gamma}^{(n)}\}\text{ to }\{\boldsymbol{\gamma}\}\\
    &\qquad\text{Set }\boldsymbol{\gamma}^{(n,g)}\text{ to the last element of }\{\boldsymbol{\gamma}^{(n)}\}\\
    &\qquad\text{Perform simulated annealing with initial guess }\boldsymbol{\gamma}^{(n,g)}\text{, and collect }\{\boldsymbol{\gamma}^{(n)}\}\\
    &\qquad\text{Set }\boldsymbol{\gamma}^{(n,g)}\text{ to the last element of }\{\boldsymbol{\gamma}^{(n)}\}\\
    &\qquad N_\text{MS}=N_\text{MS}+1\\
    &\text{End}
\end{align*}
At this point, we obtain a large collection of samples $\{\boldsymbol{\gamma}\}$. This sample is post-processed and sorted to give the 95\% credible interval:
\begin{align*}
    &\text{For each element in }\{\boldsymbol{\gamma}\},\\
    &\qquad\text{Calculate }\boldsymbol{\beta}\text{ and }\{G_l(\textbf{v})\}\\
    &\text{End}\\
    &\text{For each element in }\boldsymbol{\beta}\text{ and }\{G_l(\textbf{v})\}\text{, e.g., }R_\infty,\\
    &\qquad\text{Sort }\{R_\infty\}\text{ in ascending order}\\
    &\qquad\text{Reject the highest and lowest }2.5\%\text{ of the set}\\
    &\text{End}
\end{align*}
The first and last elements of each set corresponds to the upper and lower 95\% credible interval for the corresponding parameter or distribution value. 

\section*{References}
\bibliographystyle{IEEEtran}
\bibliography{main}

\begin{thebibliography}{10}
\providecommand{\url}[1]{#1}
\csname url@samestyle\endcsname
\providecommand{\newblock}{\relax}
\providecommand{\bibinfo}[2]{#2}
\providecommand{\BIBentrySTDinterwordspacing}{\spaceskip=0pt\relax}
\providecommand{\BIBentryALTinterwordstretchfactor}{4}
\providecommand{\BIBentryALTinterwordspacing}{\spaceskip=\fontdimen2\font plus
\BIBentryALTinterwordstretchfactor\fontdimen3\font minus
  \fontdimen4\font\relax}
\providecommand{\BIBforeignlanguage}[2]{{%
\expandafter\ifx\csname l@#1\endcsname\relax
\typeout{** WARNING: IEEEtran.bst: No hyphenation pattern has been}%
\typeout{** loaded for the language `#1'. Using the pattern for}%
\typeout{** the default language instead.}%
\else
\language=\csname l@#1\endcsname
\fi
#2}}
\providecommand{\BIBdecl}{\relax}
\BIBdecl

\bibitem{Bonora96}
P.~L. Bonora, F.~Deflorian, and L.~Fedrizzi, ``Electrochemical impedance
  spectroscopy as a tool for investigating underpaint corrosion,''
  \emph{Electrochimica Acta}, vol.~41, no. 7--8, pp. 1073--1082, 1996.

\bibitem{Juttner90}
K.~J\"{u}ttner, ``Electrochemical impedance spectroscopy (eis) of corrosion
  processes on inhomogeneous surfaces,'' \emph{Electrochimica Acta}, vol.~35,
  no.~10, pp. 1501--1508, 1990.

\bibitem{Mansfeld90}
F.~Mansfeld, ``Electrochemical impedance spectroscopy (eis) as a new tool for
  investigating methods of corrosion protection,'' \emph{Electrochimica Acta},
  vol.~35, no.~10, pp. 1533--1544, 1990.

\bibitem{Mansfeld95}
------, ``Use of electrochemical impedance spectroscopy for the study of
  corrosion protection by polymer coatings,'' \emph{Journal of Applied
  Electrochemistry}, vol.~25, no.~3, pp. 187--202, 1995.

\bibitem{Boukamp04}
B.~A. Boukamp, ``Electrochemical impedance spectroscopy in solid state ionics:
  recent advances,'' \emph{Solid State ionics}, vol. 169, no. 1--4, pp. 65--73,
  2004.

\bibitem{Fu15a}
Y.~Fu, S.~Poizeau, A.~Bertei, C.~Qi, A.~Mohanram, J.~D. Pietras, and M.~Z.
  Bazant, ``Heterogeneous electrocatalysis in porous cathodes of solid oxide
  fuel cells,'' \emph{Electrochimica Acta}, vol. 159, pp. 71--80, 2015.

\bibitem{Fu15b}
Y.~Fu, Y.~Jiang, S.~Poizeau, A.~Dutta, A.~Mohanram, J.~D. Pietras, and M.~Z.
  Bazant, ``Multicomponent gas diffusion in porous electrodes,'' \emph{Journal
  of the Electrochemical Society}, vol. 162, no.~6, pp. F613--F621, 2015.

\bibitem{He09}
Z.~He and F.~Mansfeld, ``Exploring the use of electrochemical impedance
  spectroscopy (eis) in microbial fuel cell studies,'' \emph{Energy and
  Environmental Science}, vol.~2, no.~2, pp. 215--219, 2009.

\bibitem{Schichlein02}
H.~Schichlein, A.~C. m\"{u}ller, M.~Voigts, A.~Kr\"{u}gel, and
  E.~Ivers-Tiff\'{e}e, ``Deconvolution of electrochemical impedance spectra for
  the identification of electrode reaction mechanisms in solid oxide fuel
  cells,'' \emph{Journal of Applied Electrochemistry}, vol.~32, no.~8, pp.
  875--882, 2002.

\bibitem{Andre11a}
D.~Andre, M.~Meiler, K.~Steiner, C.~Wimmer, T.~Soczka-Guth, and D.~U. Sauer,
  ``Characterization of high-power lithium-ion batteries by electrochemical
  impedance spectroscopy. i. experimental investigation,'' \emph{Journal of
  Power Sources}, vol. 196, no.~12, pp. 5334--5341, 2011.

\bibitem{Andre11b}
------, ``Characterization of high-power lithium-ion batteries by
  electrochemical impedance spectroscopy. ii: Modelling,'' \emph{Journal of
  Power Sources}, vol. 196, no.~12, pp. 5349--5356, 2011.

\bibitem{Deng13}
Z.~Deng, Z.~Zhang, Y.~Lai, J.~Liu, J.~Li, and Y.~Liu, ``Electrochemical
  impedance spectroscopy study of a lithium/sulfur battery: modeling and
  analysis of capacity fading,'' \emph{Journal of the Electrochemical Society},
  vol. 160, no.~4, pp. A553--A558, 2013.

\bibitem{Song13}
J.~Song and M.~Z. Bazant, ``Effects of nanoparticle geometry and size
  distribution on diffusion impedance of battery electrodes,'' \emph{Journal of
  the Electrochemical Society}, vol. 160, no.~1, pp. A15--A24, 2013.

\bibitem{Song14}
------, ``Electrochemical impedance of a battery electrode with anisotropic
  active particles,'' \emph{Electrochimica Acta}, vol. 131, pp. 214--227, 2014.

\bibitem{Pejcic06}
B.~Pejcic and R.~{de Marco}, ``Impedance spectroscopy: Over 35 years of
  electrochemical sensor optimization,'' \emph{Electrochimica Acta}, vol.~51,
  no.~28, pp. 6217--6229, 2006.

\bibitem{Adachi06}
M.~Adachi, M.~Sakamoto, J.~Jiu, Y.~Ogata, and S.~Isoda, ``Determination of
  parameters of electron transport in dye-sensitized solar cells using
  electrochemical impedance spectroscopy,'' \emph{Journal of Physical Chemistry
  B}, vol. 110, no.~28, pp. 13\,872--13\,880, 2006.

\bibitem{Wang05}
Q.~Wang, J.~E. Moser, and M.~Gr{\"{a}}tzel, ``Electrochemical impedance
  spectroscopic analysis of dye-sensitized solar cells,'' \emph{Journal of
  Physical Chemistry B}, vol. 109, no.~31, pp. 14\,945--14\,953, 2005.

\bibitem{Lisdat08}
F.~Lisdat and D.~Sch{\"{a}}fer, ``The use of electrochemical impedance
  spectroscopy for biosensing,'' \emph{Analytical and Bioanalytical Chemistry},
  vol. 391, no.~5, pp. 1555--1567, 2008.

\bibitem{Ruan02}
C.~Ruan, L.~Yang, and Y.~Li, ``Immunobiosensor chips for detection of
  escherichia coli o157: H7 using electrochemical impedance spectroscopy,''
  \emph{Analytical Chemistry}, vol.~74, no.~18, pp. 4814--4820, 2002.

\bibitem{Barsoukov18}
E.~Barsoukov and J.~R. Macdonald, \emph{Impedance spectroscopy: Theory,
  experiment, and applications}, 3rd~ed.\hskip 1em plus 0.5em minus 0.4em\relax
  New Jersey: John Wiley \& Sons, 2018.

\bibitem{Lasia14}
A.~Lasia, \emph{Electrochemical impedance spectroscopy and its applications},
  8th~ed.\hskip 1em plus 0.5em minus 0.4em\relax New York: Springer, 2014.

\bibitem{Orazem97}
M.~E. Orazem, J.~M. Esteban, K.~J. Kennelley, and R.~M. Degerstedt,
  ``Mathematical models for cathodic protection of an underground pipeline with
  coating holidays: Part 1 - theoretical development,'' \emph{Corrosion
  Science}, vol.~53, no.~4, pp. 264--272, 1997.

\bibitem{Song19}
J.~Song, E.~Khoo, and M.~Z. Bazant, ``Electrochemical impedance of
  electrodiffusion in charged medium under dc bias,'' \emph{arXiv}, 2019.

\bibitem{Boukamp15}
B.~A. Boukamp, ``Fourier transform distribution function of relaxation times;
  application and limitations,'' \emph{Electrochimica Acta}, vol. 154, pp.
  35--46, 2015.

\bibitem{Ciucci15}
F.~Ciucci and C.~Chen, ``Analysis of electrochemical impedance spectroscopy
  data using the distribution of relaxation times: A bayesian and hierarchical
  bayesian approach,'' \emph{Electrochimica Acta}, vol. 167, pp. 439--454,
  2015.

\bibitem{Florsch14}
N.~Florsch, A.~Revil, and C.~Camerlynck, ``Inversion of generalized relaxation
  time distributions with optimized damping parameters,'' \emph{Journal of
  Applied Geophysics}, vol. 109, pp. 119--132, 2014.

\bibitem{Hershkovitz11a}
S.~Hershkovitz, S.~Baltianski, and Y.~Tsur, ``Harnessing evolutionary
  programming for impedance spectroscopy analysis: A case study of mixed
  ionic-electronic conductor,'' \emph{Solis State Ionics}, vol. 188, no.~1, pp.
  104--109, 2011.

\bibitem{Hershkovitz11b}
S.~Hershkovitz, S.~Tomer, S.~Baltianski, and Y.~Tsur, ``Isgp: Impedance
  spectroscopy analysis using evolutionary programming procedure,'' \emph{ECS
  Transactions}, vol.~33, no.~40, pp. 67--73, 2011.

\bibitem{Horlin98}
T.~H\"{o}rlin, ``Deconvolution and maximum entropy in impedance spectroscopy of
  noninductive systems,'' \emph{Solid State Ionics}, vol. 107, no. 3--4, pp.
  241--253, 1998.

\bibitem{Saccoccio14}
M.~Saccoccio, T.~W. Han, C.~Chen, and F.~Ciucci, ``Optimal regularization in
  distribution of relaxation times applied to electrochemical impedance
  spectroscopy: Ridge and lasso regression methods - a theoretical and
  experimental study,'' \emph{Electrochimica Acta}, vol. 147, pp. 470--482,
  2014.

\bibitem{Han15}
T.~W. Han, M.~Saccoccio, C.~Chen, and F.~Ciucci, ``Influence of the
  discretization methods on the distribution of relaxation times deconvolution:
  Implementing radial basis functions with drttools,'' \emph{Electrochimica
  Acta}, vol. 184, pp. 483--499, 2015.

\bibitem{Tricomi85}
F.~G. Tricomi, \emph{Integral equations}, 1st~ed.\hskip 1em plus 0.5em minus
  0.4em\relax Massachusetts: Courier Corporation, 1985.

\bibitem{Whittaker96}
E.~T. Whittaker and G.~N. Watson, \emph{A course of modern analysis},
  1st~ed.\hskip 1em plus 0.5em minus 0.4em\relax New York: Cambridge university
  press, 1996.

\bibitem{Bazant01}
M.~Z. Bazant and B.~L. Trout, ``A method to extract potentials from the
  temperature dependence of langmuir constants for clathrate-hydrates,''
  \emph{Physica A: Statistical Mechanics and its Applications}, vol. 300, no.
  1--2, pp. 139--173, 2001.

\bibitem{Klotz12}
D.~Klotz, J.~P. Schmidt, A.~Kromp, A.~Weber, and E.~Ivers-Tiff\'{e}e, ``The
  distribution of relaxation times as beneficial tool for equivalent circuit
  modeling of fuel cells and batteries,'' \emph{ECS Transactions}, vol.~41,
  no.~28, pp. 25--33, 2012.

\bibitem{Li18}
W.~Li, Y.~Li, M.~Fang, M.~Shui, and J.~Shu, ``Distribution of relaxation times
  investigation of $\text{Co}^{3+}$ doping lithium-rich cathode material
  $\text{Li}[\text{Li}_{0.2}\text{Ni}_{0.1}\text{Mn}_{0.5}\text{Co}_{0.2}]\text{O}_2$,''
  \emph{Bulletin of Materials Science}, vol.~41, no.~6, p. 155, 2018.

\bibitem{Schmidt13}
J.~P. Schmidt, P.~Berg, M.~Sch\"{o}nleber, A.~Weber, and E.~Ivers-Tiff\'{e}e,
  ``The distribution of relaxation times as basis for generalized time-domain
  models for li-ion batteries,'' \emph{Journal of Power Sources}, vol. 221, pp.
  70--77, 2013.

\bibitem{Schindler16}
S.~Schindler, A.~Weiss, S.~Galbiati, F.~Mack, M.~A. Danzer, and R.~Zeis,
  ``Identification of polarization losses in high temperature pem fuel cells by
  distribution of relaxation times analysis,'' \emph{ECS Transactions},
  vol.~75, no.~14, pp. 45--53, 2016.

\bibitem{Sumi17}
H.~Sumi, H.~Shimada, Y.~Yamaguchi, and T.~Yamaguchi, ``Distribution of
  relaxation times analysis for optimization of anode thickness in
  metal-supported microtubular solid oxide fuel cells,'' \emph{ECS
  Transactions}, vol.~78, no.~1, pp. 2151--2157, 2017.

\bibitem{Zhang16}
Y.~Zhang, Y.~Chen, M.~Li, M.~Yan, M.~Ni, and C.~Xia, ``A high-precision
  approach to reconstruct distribution of relaxation times from electrochemical
  impedance spectroscopy,'' \emph{Journal of Power Sources}, vol. 308, pp.
  1--6, 2016.

\bibitem{Macdonald95}
J.~R. Macdonald, ``Exact and approximate nonlinear least-squares inversion of
  dielectric relaxation spectra,'' \emph{The Journal of Chemical Physics}, vol.
  102, no.~15, pp. 6241--6250, 1995.

\bibitem{Tuncer06}
E.~Tuncer and J.~R. Macdonald, ``Comparison of methods for estimating
  continuous distributions of relaxation times,'' \emph{Journal of Applied
  Physics}, vol.~99, no.~7, p. 074106, 2006.

\bibitem{Song18}
J.~Song and M.~Z. Bazant, ``Electrochemical impedance imaging via the
  distribution of diffusion times,'' \emph{Physical Review Letters}, vol. 120,
  no.~11, p. 116001, 2018.

\bibitem{Zic20}
M.~\v{Z}ic, S.~Pereverzyev, V.~Suboti\'{c}, and S.~Pereverzyev, ``Adaptive
  multi-parameter regularization approach to construct the distribution
  function of relaxation times,'' \emph{GEM - The International Journal of
  Geomathematics}, vol.~11, no.~1, p.~2, 2020.

\bibitem{Tuncer01}
E.~Tuncer and S.~M. Guba\'{n}ski, ``On dielectric data analysis using the monte
  carlo method to obtain relaxation time distribution and comparison non-linear
  spectral function fits,'' \emph{IEEE Transactions on Dielectrics and
  Electrical Insulation}, vol.~8, pp. 310--320, 2001.

\bibitem{Hansen94}
P.~C. Hansen, ``Regularization tools: a matlab package for analysis and
  solution of discrete ill-posed problems,'' \emph{Numerical Algorithms},
  vol.~6, no.~1, pp. 1--35, 1994.

\bibitem{Dion99}
F.~Dion and A.~Lasia, ``The use of regularization methods in the deconvolution
  of underlying distributions in electrochemical processes,'' \emph{Journal of
  Electroanalytical Chemistry}, vol. 475, no.~1, pp. 28--37, 1999.

\bibitem{effendy21}
S.~Effendy, Y.~R. Zhou, and M.~Bazant, ``Application of eis and machine
  learning for predicting the failure of paint-coated steel.''

\bibitem{Amin07}
M.~A. Amin, S.~S.~A. El-Rehim, E.~E.~F. El-Sherbini, and R.~S. Bayoumi, ``The
  inhibition of low carbon steel corrosion in hydrochloric acid solutions by
  succinic acid: Part i. weight loss, polarization, eis, pzc, edx and sem
  studies,'' \emph{Electrochimica Acta}, vol.~52, no.~11, pp. 3588--3600, 2007.

\bibitem{Han14}
J.-H. Han, E.~Khoo, P.~Bai, and M.~Z. Bazant, ``Over-limiting current and
  control of dendritic growth by surface conduction in nanopores,''
  \emph{Scientific Reports}, vol.~4, p. 7056, 2014.

\bibitem{Ruffo09}
R.~Ruffo, S.~S. Hong, C.~K. Chan, R.~A. Huggins, and Y.~Cui, ``Impedance
  analysis of silicon nanowire lithium ion battery anodes,'' \emph{Journal of
  Physical Chemistry C}, vol. 113, no.~26, pp. 11\,390--11\,398, 2009.

\bibitem{Balasubramanian97}
V.~Balasubramanian, ``Statistical inference, occam's razor, and statistical
  mechanics on the space of probability distributions,'' \emph{Neural
  Computation}, vol.~9, no.~2, pp. 349--368, 1997.

\bibitem{Blumer87}
A.~Blumer, A.~Ehrenfeucht, D.~Haussler, and M.~K. Warmuth, ``Occam's razor,''
  \emph{Information Processing Letters}, vol.~24, no.~6, pp. 377--380, 1987.

\bibitem{Domingos99}
P.~Domingos, ``The role of occam's razor in knowledge discovery,'' \emph{Data
  mining and Knowledge Discovery}, vol.~3, no.~4, pp. 409--425, 1999.

\bibitem{Chen90}
N.-X. Chen, Y.~Chen, and G.-Y. Li, ``Theoretical investigation on inversion for
  the phonon density of states,'' \emph{Physics Letters A}, vol. 149, no. 7--8,
  pp. 357--364, 1990.

\bibitem{Hughes90}
B.~D. Hughes, N.~E. Frankel, and B.~W. Ninham, ``Chen’s inversion formula,''
  \emph{Physical Review A}, vol.~42, no.~6, p. 3643, 1990.

\bibitem{Lakhtakia84}
M.~Lakhtakia and A.~Lakhtakia, ``Inverse black body radiation at submillimeter
  wavelengths,'' \emph{IEEE Transactions on Antennas and Propagation}, vol.~32,
  no.~8, pp. 872--873, 1984.

\bibitem{Weiss59}
G.~Weiss, ``On the inversion of the specific-heat function,'' \emph{Progress of
  Theoretical Physics}, vol.~22, no.~4, pp. 526--530, 1959.

\bibitem{Hansen93}
P.~C. Hansen and D.~P. O'Leary, ``The use of the l-curve in the regularization
  of discrete ill-posed problems,'' \emph{SIAM Journal on Scientific
  Computing}, vol.~14, no.~6, pp. 1487--1503, 1993.

\bibitem{Hansen99}
P.~C. Hansen, ``The l-curve and its use in the numerical treatment of inverse
  problems,'' in \emph{Computational inverse problems in electrocardiology},
  P.~C. Hansen and P.~Johnston, Eds.\hskip 1em plus 0.5em minus 0.4em\relax
  Southampton: WIT Press, 1999, pp. 119--142.

\bibitem{Randles47}
J.~E.~B. Randles, ``Kinetics of rapid electrode reactions,'' \emph{Discussions
  of the Faraday Society}, vol.~1, pp. 11--19, 1947.

\bibitem{Deb00}
K.~Deb, S.~Agrawal, A.~Pratap, and T.~Meyarivan, ``A fast elitist non-dominated
  sorting genetic algorithm for multi-objective optimization: Nsga-ii,'' in
  \emph{International conference on parallel problem solving from nature}, New
  York, 2000, pp. 849--858.

\end{thebibliography}

\end{document}